\documentclass[acmsmall]{acmart}

\AtBeginDocument{%
  \providecommand\BibTeX{{%
    \normalfont B\kern-0.5em{\scshape i\kern-0.25em b}\kern-0.8em\TeX}}}

\setcopyright{acmcopyright}
\copyrightyear{2022}
\acmYear{2022}
\acmDOI{XXXXXXX.XXXXXXX}
\acmJournal{CSUR}
    
\usepackage{array}
\usepackage{varwidth}
\usepackage{longtable}
\usepackage{float}
\usepackage{color, colortbl}
\usepackage{graphicx}
\usepackage{makecell}

\definecolor{Gray}{gray}{0.9}

\begin{document}

\title[When Industry meets Trustworthy AI]{When Industry meets Trustworthy AI: A Systematic Review of AI for Industry 5.0}
\author{Eduardo Vyhmeister}
\authornote{Both authors contributed equally to this research.}
\email{eduardo.vyhmeister@insight-centre.org}

\author{Gabriel G. Castañe}
\authornotemark[1]
\email{gabriel.gonzalezcastane@ucc.ie}

\affiliation{%
  \institution{University College Cork}
  \streetaddress{Western Gateway Building, University College Cork, Western Road}
  \city{Cork}
  \state{Cork}
  \country{Ireland}
  \postcode{T12YN60}
}

\renewcommand{\shortauthors}{Castañé and Vyhmeister}

\begin{abstract}
Industry is at the forefront of adopting new technologies, and the process followed by the adoption has a significant impact on the economy and society. In this work, we focus on analysing the current paradigm in which industry evolves, making it more sustainable and Trustworthy. In Industry 5.0, Artificial Intelligence (AI), among other technology enablers, is used to build services from a sustainable, human-centric and resilient perspective. It is crucial to understand those aspects that can bring AI to industry, respecting Trustworthy principles by collecting information to define how it is incorporated in the early stages, its impact, and the trends observed in the field. In addition, to understand the challenges and gaps in the transition from Industry 4.0 to Industry 5.0, a general perspective on the industry's readiness for new technologies is described. This provides practitioners with novel opportunities to be explored in pursuit of the adoption of Trustworthy AI in the sector. 
\end{abstract}



\maketitle
\section{Introduction}
Industrial processes and factories are evolving to transition to digitalization with novel techniques that automate data computations more efficiently -- and adequately -- to improve tasks. The fourth Industrial Revolution or Industry 4.0 incorporated cyber-physical systems into factories within production systems, warehousing, and logistics, easing the adoption of new digital assets and establishing them within enterprise values \cite{frazzon_towards_2013}. In this transition process, the decrease of the hardware costs, and the vast amounts of data to be processed impulsed novel techniques that automates, secures, and flexibly adapt factories. Foremost among these are: AI to boost data analysis and outcome predictions \cite{european_commission_joint_research_centre_ai_2021}; Blockchain to secure and transparent record and share data \cite{alves_towards_2022}; Additive Manufacturing to rapidly prototype and customise the products \cite{amini_cybermanufacturing_2019}; Internet of the Things (IoT), Edge and Big Data analyses to enable real-time data processing and analysis for decision making \cite{vermesan_next_2018}. 

The industrial ecosystem demands a transition to embrace resilient sources that promote societal and environmental wellness while driving economic growth. Although this can seem as utopian, the European Commission has designed a strategic path that, effectively executed, will transforms traditional factories into resilient providers of prosperity, thereby evolving production centres into respectful components for environmental and societal well-being~\cite{european-commission-industry-2020}. This path is a transformative journey that defines the development from Industry 4.0 into a new industrial revolution, Industry 5.0~\cite{brunetti_smart_2022}.

Between all enabling technologies with potential for growth from Industry 4.0 to Industry 5.0, Artificial Intelligence (AI) is widely recognized as a key among them. The automation or semi-automation of large-scale production processes and extensive customizations makes it invaluable in this transition. However, the use of AI must be underpinned by a high degree of trustworthiness in the use of AI to bootstrap a technological transition. Hence, the economic benefits became a secondary dimension compared to the minimization of associated risks and their impacts on society and workers in factories~\cite{pieters_explanation_2011,quinn_trust_2021,devitt_trust_2021}. This makes the adoption of AI technology an important case of study encompassing socio-technical and legal challenges that must be addressed to reach human and machine cooperation in smart-working environments~\cite{bednar_socio-technical_2020}.

However, factories are struggling to adopt operations to the standards of Industry 4.0. There are several challenges that hamper a successful integration of necessary technologies for transitioning to this stage \cite{elibal_systematic_2021,rakic_precondition_2021}. In addition, the emergence of Industry 5.0 can be perceived as an extra layer of complexity for manufacturers and factories to evolve in the digitalization process \cite{fujimaki_6_2020}.
Despite that some production centres are not ready for the adoption, early studies show that transitioning to Industry 5.0 pose benefits for companies such as: a seamless adoption of their products in the market by making them trustworthy and explainable, advantages in product assortments \cite{lauer_you_2021}, improvement in maintenance processes and cost reduction, and enhancement in product quality \cite{brosset_scaling_nodate}.

From a legal perspective, governments are working on several AI regulations. 
For example, the AI Act \cite{high-level_expert_group_on_artificial_intelligence_ethics_2019,european_commission_laying_2021} creates awareness among companies to promote the adoption of Industry 5.0. 
The major challenge for industry is to identify the correct paths and challenges within the extensive work provided by the research community on Trustworthy AI topics. The diverse literature on each of these challenges -- when/if identified -- hampers their understanding of the necessary steps to meet the necessary requirements.

The work here presented provides a comprehensive review of the literature on state-of-the-art Trustworthy AI technologies used in industry and the opportunities for transitioning to Industry 5.0. The analysis explores research trends and highlights critical aspects of AI adoption by the industrial sector. While various studies have examined the status of Industry 4.0, including bibliometric evaluations analyzing trends and gaps \cite{muhuri_industry_2019,ahmi_bibliometric_2019,liao_past_2017,kipper_scopus_2020}, different research approaches and statuses used in Industry 4.0 \cite{kamble_sustainable_2018,ali_industry_2022}, future trends and challenges \cite{ali_industry_2022,elibal_systematic_2021}, and the impact of Industry 4.0 on business models \cite{grabowska_business_2022}, to the best of the authors' knowledge, no articles exist that define the state-of-the-art dimensions of Trustworthy AI comprehensively to be applied in Industry 4.0 and, consequently, to reach Industry 5.0. The data for this study was collected through a systematic literature review of peer-reviewed articles to analyze the trends.

The paper is organized as follows:
Section \ref{background} provides an overview of the main aspects related to Industry 4.0 and Industry 5.0, along with the leading AI technologies applied in these contexts.
Section \ref{research methodology} describes the research methodology and materials used for the literature review, including qualitative and quantitative results for each research question.
Section \ref{discussion} presents a summary of the discussion and findings.
Finally, Section \ref{conclusions} presents the study's conclusions and offers a general perspective on the observed trends.

\section{Background} \label{background}
This section provides an overview of Industry 4.0 and Industry 5.0 to give the reader a comprehensive understanding of the existing challenges. In addition, it also describes the topics where AI is applied to the industry are also described.

\subsection{Industry 4.0 and Industry 5.0}
Over the last decade, Industry 4.0 focused on incorporating cyber-physical systems and digital platforms into factories, integrating monitoring processes~\cite{zheng_applications_2021}. 
By harnessing real-time data, factories are able to gather and analyze valuable information, thus providing valuable insights for decision-making processes. This has led companies in two directions to enhance their production capabilities: decision making features and decentralization of production processes from the traditional production control systems~\cite{ferber_industry_2012}.

In the scope of decision-making, Industry 4.0 explores the use of technological advancements for training and decision-making using models that replicate real-world situations in controlled and safe environments, such as Digital Twins and Augmented Reality acting to facilitate (enable) enhanced working environments and experiences in industrial settings, integrating virtual and real systems, both independently and collaboratively with human counterparts. As a consequence, this approach enables a thorough examination of business objective decisions while prioritizing risk mitigation, leading to the minimization of adverse economic consequences on factories and products.

The decentralization has been gathered by sensors and actuators operating on different machines through IoT, facilitating comprehensive connectivity with computing systems and resulting in the generation of large volumes of streaming data, commonly known as Big Data~\cite{misra_iot_2022}. 
The processing of this data is carried out using IoT devices either locally or on the Cloud/Edge, which reduces costs and enhances scalability through the utilization of virtual resources in the Cloud or at the network Edge ~\cite{zheng_applications_2021}. 
The increased computational power and reduction of costs, enhanced the ability to process data efficiently, enabling the adoption of multiple technologies that support the goals of Industry 4.0. However, to manage such vast amounts of data, novel methods for intelligent acquisition, collection, and process are required~\cite{xu_big_2019}. Effective data management and processing techniques contribute to improved scalability, security, and efficiency but also a reduction in the required resources. This is the case for Additive Manufacturing that streamlines the assembly process by reducing the number of critical components required \cite{lennon_olsen_industry_2019}. 
Although AI is not officially considered a fundamental pillar of Industry 4.0, novel techniques supported by AI have a significant role in driving its advancement. Numerous companies have developed innovative intelligent systems enabling a certain level of process automation. 

By utilizing AI, the information from various systems can be efficiently gathered and processed at high speed, enabling the execution of tasks, such as Fault Prediction and Action Selection \cite{angelopoulos_tackling_2019}. AI enables intelligent decision-making and predictive capabilities, allowing for more efficient utilization of the gathered data in Industry 4.0 systems. In summary, AI plays a crucial role in leveraging IoT services and Cloud Computing to enhance the enablers and technologies for industry. 

Table \ref{table:industry4_facet} shows a list of the most relevant pillars that drive Industry 4.0 as extracted from \cite{silvestri_maintenance_2020}. 

\begin{table}[]
\caption{Industry 4.0 Technological Enablers}
\label{table:industry4_facet}
    \begin{centering}
     \centering{}
    \small\addtolength{\tabcolsep}{0pt}
        \begin{tabular}{|p{2cm}|p{11cm}|}
        \hline \rowcolor{Gray}
         Category & Description   \tabularnewline
        \hline 
        \makecell[tl]{Additive \\ Manufacturing} & Also known as 3D printing, encompasses a range of techniques dedicated to the production of products through layer-by-layer deposition \cite{zheng_applications_2021}. These methods include vat photo-polymerization, powder bed fusion, binder jetting, material jetting, sheet lamination, material extrusion, contour crafting, cellular fabrication, d-shape, concrete printing, and direct energy deposition \cite{butt_strategic_2020}. The benefits include reduced costs, reduced supply chain, worker safety, complex forms fabrication, and short turnaround times.%
        \tabularnewline  \hline
        
        \makecell[tl]{Augmented \\ Reality} & By overlaying digital content onto the physical environment, this technology serves as a bridge between the digital and real-world \cite{butt_strategic_2020}. It offers diverse benefits, including improved product development insights, enhanced maintenance, training, issue resolution, support, quality assurance, and automation \cite{butt_strategic_2020}. AR utilizes markers, holograms, mobile devices, tracking, and interaction methods to enable real-time information streaming. This facilitates the monitoring and control of virtual representations known as Digital Twins, enhancing the management of industrial processes \cite{butt_strategic_2020}..
        \tabularnewline\hline
        
        Simulation & Mathematical representations of systems and phenomena offer approaches that enable the evaluation of different alternatives or scenarios within a simulated environment. These approaches find extensive application in investment assessment, production planning, optimization and scheduling, design, capability planning, process improvement, bottleneck analysis, and resource allocation. They serve as powerful tools for decision-making and analysis, providing insights and facilitating informed choices within complex systems and dynamic environments.
        \tabularnewline\hline
        
        Autonomous Systems & These systems exhibit a level of autonomy as they perceive and respond to external information, gathered through various sensors, thus demonstrating a form of intelligence \cite{oztemel_literature_2020}. They are capable of executing repetitive, hazardous, and time-consuming tasks with high accuracy and efficiency, without the need for frequent interruptions \cite{butt_strategic_2020}.
        \tabularnewline\hline
        
        \makecell[tl]{Internet of \\ Things } & The interconnection of machinery and sensors through communication channels, such as the internet, provides the foundation for IoT. In the industrial sector, this technology brings various advantages, including cost reduction, mass customization, improved safety, and accelerated time to market. Extensive literature coverage underscores the significance of IoT as a crucial enabler for other Industry 4.0 advancements \cite{group_internet_2015,laghari_review_2022}. It plays a pivotal role in driving the transformation and progress within the Industry 4.0 landscape.
        \tabularnewline \hline
        
        \makecell[tl]{Big Data and \\ Analytics} & Aggregated data's significance in the industry became apparent as it emerged as a valuable tool for decision-making processes \cite{butt_strategic_2020}. Data utilisation is further enhanced by leveraging supporting technologies, such as AI, which can analyze and harness more extensive datasets. The accumulated information is commonly employed for descriptive, exploratory, predictive, and prescriptive tasks \cite{fosso_wamba_how_2015}. \tabularnewline \hline
        
        \makecell[tl]{Cloud \\ Computing} & The utilization of computational resources through the internet, as facilitated by Cloud Computing, is strongly intertwined with IoT and has significant implications in the manufacturing sector, as evidenced in the literature \cite{novais_systematic_2019,askary_Cloud_2020}. Cloud Computing's broad impact drives efficiency, scalability, and digital transformation across various industries.
        \tabularnewline \hline
        
        Cybersecurity & Securing data systems in cyber-physical systems is crucial. This area involves policies and practices to prevent attacks and unauthorized access, meets manufacturer and consumer requirements. It ensures system integrity and protects against threats and vulnerabilities.
        \tabularnewline \hline
        
        \makecell[tl]{Horizontal and \\ Vertical \\ Integration} & Protocols and approaches defining machine and customer integration within the production system encompass Horizontal and Vertical Integration. These strategies involve acquiring related businesses or controlling production/distribution stages, respectively, to consolidate market position and differentiate from competitors \cite{peres_idarts_2018}. They expand the traditional perspective of product-to-service integration and contribute to a comprehensive industrial process.
        \tabularnewline
        \hline
        \end{tabular}
    \par\end{centering}
\end{table}

The impact of the aforementioned technologies extends beyond the industrial sector, encompassing home products, business models, clean energy, and broader sustainable aspects that were not fully taken into account in previous industrial revolutions. Moreover, the industry is recognized to be a catalyst for systemic transformation towards more sustainable economies \cite{union_industry_2022}. Therefore, it is essential to incorporate various societal and environmental aspects as new driving forces within the industrial sector.

Industry 4.0 does not fully meet Europe's objectives as it gives priority to economic-driven business optimization, which could lead to technological monopolies and increased inequalities \cite{union_industry_2022}. For example, it lacks a comprehensive framework to promote circular economies and restorative feedback loops within value chains; priorities for well-being as a fundamental pillar by considering human capabilities both within and outside industrial environments (e.g., through social governance); and a strong emphasis on the environmental impact by focusing on efficiency improvements and the adoption of clean energy sources.

Hence, in 2020, the concept of Industry 5.0 was introduced during a workshop organized by the European Commission, where research and technology organizations and funding agencies discussed the future vision for industry. This new concept incorporated AI and the societal dimension as catalysts for the future roadmap for European industry \cite{european_commission_directorate_general_for_research_and_innovation_enabling_2020}.

An early definition of Industry 5.0 is provided in the document:
\newline

``\textit{Industry 5.0 recognises the power of industry to achieve societal goals beyond jobs and growth to become a provider of prosperity by making production respect the boundaries of our planet and placing the well-being of the industry worker at the centre of the production process.}.''
\newline

Several initiatives have been designed since then to support Industry 5.0, including efforts on up-skilling and re-skilling European workers, especially in digital skills - \textit{Skills Agenda} and \textit{}{Digital Education Action Plan} \cite{european2020european}; a more competitive industry fostered by speeding up investment in research and innovation - \textit{Industrial Strategy}; advancement on sustainable development, which is translated into resource efficient and sustainable industries environment-friendly that lead to a transition towards a circular economy - \textit{Green Deal} \cite{puaschunder_legal_2019, claeys2019make}; and the adoption of a human-centric approach for digital technologies, defined through different proposals for AI regulation - including the AI Act, the white paper, and the Trustworthy AI requirements \cite{high-level_expert_group_on_artificial_intelligence_ethics_2019,european_commission_regulation_2021,european_parliament_directorate_general_for_internal_policies_of_the_union_white_2020}. 

A core pillar of the Industry 5.0 is the adoption of AI. The process targets several categories: processing data at speed, skilled employees to operate heterogeneous technologies – AI, computing resources, data – and the incorporation of ethical dimensions in the AI life-cycle to generate trust and ensure safe working environments \cite{european_commission_laying_2021, european_commission_directorate_general_for_communications_networks_content_and_technology_ethics_2019}.

However, despite the efforts of researchers to incorporate ethical considerations in AI applications at large, the AI technology used in operation environments, and the application area where these technologies are deployed, introduce unique concerns. In the case of Industry 5.0, the connection between the factory's challenges along with the AI application domains and technological pillars, remains unclear \cite{ryan_social_2022}.

\subsection{Towards Industry 5.0}

Industry 5.0 represents a paradigm shift that goes beyond solely technological and economic considerations in the industrial domain. It emphasizes the integration of human progress and well-being, encompassing sustainable, circular, and regenerative economies. This evolution acknowledges the symbiotic interaction between machines and humans, leveraging the strengths of each counterpart instead of pursuing replacement \cite{brunetti_smart_2022}. It envisions a collaborative environment where robots and technologies are designed with a human-centric approach, aiming to optimize the processes in the factories and the performance of workers.

There are different technologies that are defined as technology enablers listed by authors at \cite{xu_industry_2021,union_enabling_2020}. These are summarised in Table \ref{table:industry5_facet}, describing each of these as core components associated with Industry 5.0.

\begin{table}[htb!]
\caption{Industry 5.0 Technology Enablers}
\label{table:industry5_facet}
    \begin{centering}
    \small\addtolength{\tabcolsep}{0pt}
        \begin{tabular}{|p{2.2cm}|p{10.8cm}|}
        \hline 
        \rowcolor{Gray}
         \textbf{Category} & \textbf{Description}   \tabularnewline
        \hline 
        \makecell[tl]{Human - \\ Machine \\ interaction} & This area involves concepts that involve collaborative work -- Cobots --, multi-lingual speech, gesture recognition and human intention prediction, AR in interactive purposes (e.g. training and inclusiveness), enhancing human physical and cognitive capabilities.     \tabularnewline \hline
        
        \makecell[tl]{Bio-inspired \\ and smart \\ materials} & The technologies focus on materials that include properties such as self-repairing, recyclables, and embedded sensor technologies with intrinsic traceability. They can be integrated within living organisms/materials, among others. \tabularnewline \hline
        
        \makecell[tl]{Digital Twins \\ and \\Simulation} & Higher-level models corresponding to digital representations of physical objects or processes. 
        The models are linked to the real environment through the sensors' information. These can support complex processes, such as autonomic resource reconfiguration, replacement, or movement, in factories, increasing the flexibility of the resources. \tabularnewline \hline
        
        \makecell[tl]{Data \\ sharing and \\ processing } & The novelty in this area is related to energy-efficient and secure data management. In addition, given the impact of AI on mining data, Big Data analytic techniques are considerably improved, becoming a new hot spot \cite{wang2022big}. \tabularnewline \hline
        
        \makecell[tl]{Artificial \\ Intelligence} & Automation and classification tasks that include, among others, causality-based AI, AutoML for adaptability without human intervention, human-centric AI, improved AI components through the use of oracles/expert knowledge, improvements on  the energy consumption of AI, and scaling capabilities in dynamic systems.   \tabularnewline \hline
        
        \makecell[tl]{Technologies \\ for energy \\ efficiency, \\ renewable, \\ storage, \\ and autonomy} & These technologies develop and operationalise approaches to achieve carbon neutrality. Foremost among these are included renewable energy sources, hydrogen and hydrogen carriers, low data transmission and data analyses. The European Commission has also involved initiatives to foster the sound development of these technologies -- Green Deal \cite{claeys2019make}. This initiative does not only include industrial concepts (e.g. healthy and affordable food). Nevertheless, it clearly states the need for the industry to foster cleaner energy and cutting-edge clean technological innovation.     \tabularnewline \hline


        \makecell[tl]{Reconfigurability} & Reconfigurability is crucial in manufacturing systems, allowing swift adaptation to dynamic market demands. It involves dynamic management of physical, informational, and organizational elements to achieve flexibility and agility in the production process. This capability addresses challenges posed by changing consumer preferences, market trends, and the need for rapid product development. It integrates technologies like IoT, AI, and Additive Manufacturing enabling dynamic customization, rapid prototyping, and mass customization strategies while promoting sustainability and environmental responsibility by reducing waste and energy consumption.
        \tabularnewline
         
        \hline
        \end{tabular}
    \par\end{centering}
\end{table}

These technologies show the inclusion in Industry of new forms of sustainable, circular and regenerative economic value creation and equitable prosperity. It is expected that future industries will have a broader impact on societal goals and contribute to a greener ecosystem~\cite{eu2022industry}.

\subsection{AI, Trustworthy AI and its link to Industry 5.0}

AI technologies are widely employed in various digital services, within prominent technological leading companies such as Google, Amazon, Apple, Facebook, and Microsoft. These companies leverage AI to power a range of services, including e-commerce, online advertising, media streaming, smart home systems, self-driving cars, and social networking. The success of these services is attributed to the convergence of media and information technology, which fosters their growth \cite{smyrnaios_leffet_2016}. Within these companies, AI is utilized for diverse purposes. It enables the creation of virtual representations of the real world, converts photos into 3D images, generates interactive maps and immersive views, enhances personal assistants, improves language recognition, and offers a wide array of Edge Computing-based services like predictive maintenance for device management and Augmented Reality experiences \cite{tazrout_gafam_2020}.

Table \ref{table:AI_in manufacturing} presents an overview of AI applications in industrial fields classified based on the research conducted by \cite{fahle_systematic_2020-1,butt_strategic_2020}.

{
\begin{table}[h!]
\caption{AI Application Methods in Industrial Environment}
\label{table:AI_in manufacturing}
    \begin{centering}
     \centering{}
    \small\addtolength{\tabcolsep}{0pt}
        \begin{tabular}{|p{3cm}|p{10cm}|}
        \hline 
        \rowcolor{Gray}
         \textbf{Category} & \textbf{Description}  \tabularnewline
        \hline 
        
        \makecell[tl]{Process planning} & Research in this field is linked to the scheduling problem in the manufacturing sector. Different approaches such as Q-learning, RF, and decision trees have been applied for this task \cite{loyer_comparison_2016} for cost prediction, energy and resource efficiency, workers localisation, and load forecasting, among other tasks.  \tabularnewline \hline
        
        \makecell[tl]{Quality control} & This research can be seen as the root of product and process quality control (which adds process control and predictive maintenance considerations). Nevertheless, reference settle a distinction by linking quality control to implementations related to the problem of reducing quality assurance costs (e.g. quality detection - CNN, SVM \cite{ma_blister_2019}; Root cause analyses \cite{lokrantz_root_2018} and classification \cite{sumesh_use_2015} - bayesian network,decision tree).  \tabularnewline \hline
        
        \makecell[tl]{Predictive maintenance} & The research conducted in this field is dedicated to estimating the valuable lifetime of parts and components. Literature is broad with using different AI techniques for regression and classification \cite{zonta_predictive_2020}.  \tabularnewline \hline
        
        \makecell[tl]{Logistics} & This area of study shares some commonalities with the scheduling problem. Both consider the efficient use, flow, and storage of material (i.e. optimisation). Nevertheless, logistics considers the products from their origin, while scheduling refers to the internal transformation of raw materials into products. Based on these, some strategies are employed for system representation in scheduling trends in the field of logistics \cite{chimunhu_review_2022}.
        \tabularnewline \hline
               
        \makecell[tl]{Assistance and \\ learning systems} & It focuses on supporting and enhancing employees' capabilities. Two key aspects are referenced to consider the assistance; guidance of individual learning processes and the control of competences saturation \cite{ullrich_assistance-_2015,mazziotti_robust_2015}. Furthermore, this cluster defines research dedicated to training concepts in manufacturing environments.\tabularnewline \hline

        \makecell[tl]{Robotics} & The research is dedicated to incorporating AI within robots. The application of Machine Learning (ML) components for automation and human collaboration is wide and includes motion, object, and human recognition, path planning, and improvement (i.e. optimisation) of automation tasks.  \tabularnewline \hline

        \makecell[tl]{Process control and \\ optimisation} & It is dedicated to implementing AI in a short-time response to systems monitoring and modification by plant-wide and individual unit real-time optimisation, parameter estimation, supervisory control, data reconciliation, alarm management, emergency shutdown, and sensor and actuator validation, among other tasks. \cite{nof_process_2009, mozaffar_mechanistic_2022}  \tabularnewline
        \hline

        \end{tabular}
    \par\end{centering}
\end{table}
}

A key aspect is trust in technology. Building trust is needed for consumer confidence across all market segments. To achieve this goal, the European Commission has prioritized Trustworthy AI, aiming to establish a framework that fosters trust in the development and adoption of AI technology. 
A first step on this are the Trustworthy AI requirements, based on 84 ethical guidelines, highlighting principles, such as transparency, justice and fairness, non-maleficence, responsibility, and privacy \cite{jobin_global_2019}. These requirements are defined in seven pillars, summarized in Table \ref{table:Trustworthy requirements}.

\begin{table}[h!]
    \caption{Trustworthy AI Requirements}
    \label{table:Trustworthy requirements}
    \begin{centering}
     \centering{}
    \small\addtolength{\tabcolsep}{0pt}
    \begin{tabular}{|p{2.5cm}|p{10.5cm}|}
        \hline 
        \rowcolor{Gray}
         \textbf{Category} & \textbf{Description}   \tabularnewline
        \hline 
        \makecell[tl]{Human Agency \\ and Oversight} & Based on the principles of human autonomy and decision-making, this requirement includes fundamental rights and agency. It links to human-centric considerations and Accountability. Approaches like Human-in-the-loop (HITL), Human-on-the-loop (HOTL), and Human-in-command (HIC) extend AI autonomy and collaboration with humans, reinforcing Human Agency and Oversight. However, HITL feasibility depends on system granularity and functionality, limiting real-time human participation in dynamic processes. Human oversight alone doesn't address dependency issues and potential overconfidence in algorithms \cite{leprince-ringuet_ais_2020}.
        \tabularnewline \hline
        
        \makecell[tl]{Technical \\ Robustness and \\ Safety} & Technical robustness and safety are vital in the manufacturing sector, addressing resilience, security, fallback plans, safety, accuracy, reliability, and reproducibility \cite{high-level_expert_group_on_artificial_intelligence_ethics_2019}. Manufacturing processes aim to withstand variability in various factors. However, the extent of robustness considerations may vary depending on the level of risk involved.\tabularnewline \hline
        
        \makecell[tl]{Privacy and \\ Data Governance} & This requirement includes privacy, quality and integrity of data, and access to data. Regulatory needs regarding data managing and handling have already been set (e.g. the GDPR regulations and \cite{high-level_expert_group_on_artificial_intelligence_ethics_2019}). The data protection regulations set rules for businesses and organisations and, simultaneously, set rights for citizens regarding their data rights and redress. \tabularnewline \hline
        
        \makecell[tl]{Transparency} & This requirement encompasses AI considerations, including explainability, as well as management considerations, such as retraceability and communication. Explainability involves the ability to inspect and replicate the decision-making mechanisms employed by AI systems, while establishing clear responsibilities and ensuring Accountability for the outcomes is essential.  \tabularnewline \hline
        
        \makecell[tl]{Diversity, \\ Non-discrimination \\ and Fairness} & Mitigating unfair bias, ensuring accessibility, universal design, and promoting stakeholder participation are key aspects of this requirement \cite{higgins_cochrane_2020}. Bias, which involves systematic errors or deviations from truth, is inherent to human nature and can be observed in social information and analyses. It is essential to recognize that AI systems may perpetuate biases present in complex historical data. Addressing bias is crucial to prevent discrimination and unfairness, emphasizing the importance of trustworthiness considerations in AI that prioritize the human factor, agents, and interactions.\tabularnewline \hline
        
        \makecell[tl]{Environmental and \\ Societal well-being} & This requirement encompasses sustainability, environmental friendliness, social impact, and the principles of a democratic society. According to \cite{chui_applying_2018}, AI offers at least 18 capabilities that can benefit the community. However, the scale of implementation determines the impact and associated risks of AI. While these considerations have long-term implications for society, there is currently no established process for managing and ensuring sustainable protocols in this context. It is therefore essential for the manufacturing sector to incorporate societal well-being into its policies, even if not explicitly stated in corporate social responsibilities. \tabularnewline \hline
        
        \makecell[tl]{Accountability} & This requirement minimizes negative impact, considers trade-offs, and ensures redress. Accountability is crucial among AI developers, consumers, and users. Clear definitions of uncertainties and responsibilities promote accountable behaviour by AI developers. In the manufacturing sector, Accountability is weakly linked to agents and their interactions. It is especially important for high-risk AI elements and establishing legal obligations. Incorporating AI assets into business products involves multiple stakeholders. \tabularnewline
        \hline
    \end{tabular}
    \par\end{centering}
\end{table}

Furthermore, Human-centric AI emphasizes the collaboration and interaction between AI assets and human agents. Algorithms can be continuously updated to consider the state, needs, experiences, and physical interactions with humans. To achieve this, a combination of sensed and historical data is used to extract patterns, choices, and other trends in their behaviour.
For an AI component to be considered human-centered, it should be explainable, verifiable, physical, collaborative, and integrative. Explainability and verifiability are linked to properties such as reliability, safety, availability, confidentiality, integrity, and maintainability. The human-centric approach to AI aligns in many directions with the Trustworthy AI requirements as defined by the European Commission's and upcoming regulations on AI.

In the context of Industry 5.0, a digitalisation strategy should consider both the enterprise's interests along the Trustworthy AI pillars. For sustainable development, economic, societal, and environmental aspects are crucial for its adoption. Additionally, the resilience of the European industry, especially in the face of disruptions due to unexpected events like a pandemic, requires strategies that focus on adaptable production capacity, flexible business processes, and robust value chains. In this context, AI technologies can play a significant role in enabling industrial adaptability, supporting these objectives.

In summary, the Trustworthy AI requirements serve as essential definitions for AI systems in transitioning from Industry 4.0 to Industry 5.0, ensuring the development of an ecosystem of trust and sustainability required for a new industrial revolution.

\section{Research methodology}\label{research methodology}

\begin{figure*}[htbp!]
    \centering
    \includegraphics[width=130mm]{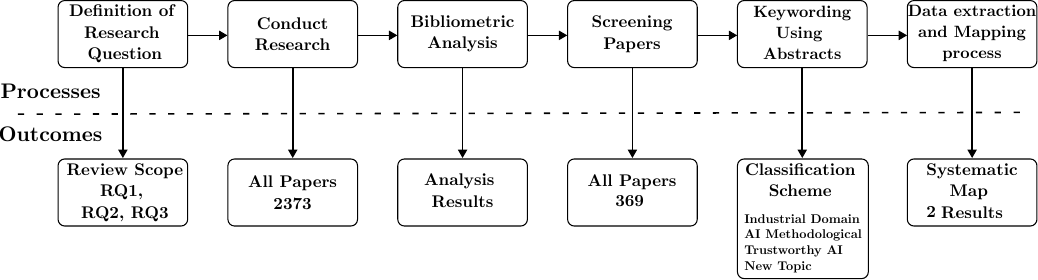}
    \caption{Systematic Mapping Approach}
    \label{fig:systematicMapping}
\end{figure*}

Figure \ref{fig:systematicMapping} shows a schematic in which the research methodology is described. The processes, on the top row of the Figure, are linked to their primary outcomes, on the bottom row in the Figure. These are review scope, all papers, bibliometric analysis, relevant papers, classification scheme, and systematic map results. Although the Bibliometric Analysis is not the core of the methodology, it is added to ground a better understanding of the topic trends. These are developed in the subsequent Subsections according to the Figure.


\subsection{Definition of Research Question}

The goal is defined as follows:\newline

``\textbf{To provide an understanding of trends and status on the implementation of human-centric and Trustworthy AI technologies in the industrial sector}''. 
\newline

From the goal, the research questions are the following:

\begin{itemize}

   \item RQ1 - What are the key Trustworthy AI requirements prioritized in Industry 4.0 technological enablers and their AI applications, and how can they be further promoted?
    
    \item RQ2 - What are the current trends in the relationship between Industry 4.0 enablers and AI applications?
    
    \item RQ3 - What innovative approaches can facilitate the integration of Trustworthy AI into Industry 4.0 to transition to Industry 5.0?

\end{itemize}

\subsection{Search Methodology}

To assess the status of AI in Industry 4.0 and establish the foundations for Industry 5.0, a mapping process and a systematic approach were conducted. The process starts with the definition of the key pillars that drive the topic and the use of targeted keywords to conduct literature searches \cite{petersen2008systematic}. These keywords were used as markers, enabling efficient search across various platforms. Keywords and pillars were combined to address the research questions. Specifically, keyword selection was approached by establishing specific groups to help with the definition of the components of the keywords, taking a bottom-up approach. The methodology applied to construct queries and definition of keywords follow the PICOC methodology -- Population, Intervention, Comparison, Outcomes, and Context \cite{petersen2015guidelines, petticrew2008systematic, keele2007guidelines}. Each of these topics is defined next:

\begin{itemize}

\item Population: It refers to the specific group of individuals or subjects under the interest of the study. In the context of this work, the population is the group of manuscripts from which research questions are formulated.
    
\item Intervention: The intervention refers to the approach or technique applied in the empirical study. This study involves software methodologies, tools, technologies, or procedures. Different AI techniques that can be applied to specific procedures in the industrial sector are considered. It is represented by the \textbf{AI Application Methods} set, which includes AI applied in Industry 4.0 (Table \ref{table:AI_in manufacturing}), and the \textbf{Trustworthy AI} set, and that defines the requirements from the Trustworthy AI guidelines (Table \ref{table:Trustworthy requirements}).

\item Comparison: The comparison component involves differentiating methods, processes, or strategies. There are different AI techniques that can be applied to specific procedures in the industrial sector, leading by a specific implementation of the EC initiatives to Industry 5.0 -- i.e. one set named \textbf{AI Application Methods} that is linked to the AI applied in Industry 4.0 (Table \ref{table:AI_in manufacturing}), and another set named \textbf{Trustworthy AI} that defines the requirements from the Trustworthy AI guidelines (Table \ref{table:Trustworthy requirements}). 

\item Outcomes: Since empirical approaches or comparisons are not considered, no specific outcomes are defined in this component.

\item Context: The context provides a comprehensive view of whether the study is conducted in academia or industry, the industrial segment, and the subject's incentives. Thus, the technological context imposes keywords to constrain the search, for example, the use of the word AI or Artificial Intelligence.
\end{itemize}

The keywords are grouped into three categories: Industry, AI Application methods, and Trustworthy AI. 

For the first category, \textit{Industry 4.0}, these are: Additive Manufacturing, Augmented Reality, Digital Twin, Autonomous Robots, Robotics, IoT, Big Data, Cloud Computing, Cybersecurity, Horizontal and Vertical Integration. 
For the second category, \textit{AI Application methods}: Process planning, Quality Control, Predictive maintenance, Logistics, Robotics, Learning systems, Process Control, or Process Optimization. 
Finally, for the third category, \textit{Trustworthy AI requirements}: Responsible AI, Trustworthy AI, Transparency, Explainability, Traceability, Human Agency, Human Oversight, Privacy, Data Governance, Accountability, Fairness, Technical Robustness, and Safety. Additionally, the list of keywords is completed with the terms AI and Artificial Intelligence, included as population components.

The search engines used for the queries are IEEE, Scopus, ACM, and Google Scholar (GS). The first three have a common search interface for their database while in this case Google Scholar is limited to tuples to perform logic queries. 

The queries on IEEE, Scopus, and ACM are designed to focus on gathering the Industry 5.0 coverage or Industry 4.0 with enablers that allow its transition to Industry 5.0.

The Google Scholar queries are structured as follows: $GS1$ is used to examine the correlation between industrial topics and on Responsible AI or Trustworthy AI; $GS2$ is designed to understand the global trend of Industry 4.0, and it is a superset of $GS3$ to $GS11$; and finally, $GS_3$ to $GS_11$ focus are on specific aspects of Industry 4.0 with AI and Manufacturing.

In the case of $GS4$, the query has been split into two queries. One is for Simulation or modelling in combination with AI, while the second targets AI applications in the context of Digital Twins. 

The time-frame for the queries is from 2012 to June 2022 to capture temporal trends. Only English publications were considered, and the analysis focused on journals, standards, and peer-reviewed conference publications.

The queries using the keywords above mentioned are shown in Table \ref{table:search}.

\begin{small}
\begin{longtable}{|p{1.5cm}|p{11.5cm}|}
\caption{Queries to the databases IEEE, Scopus, ACM and Google Scholar}
\label{table:search}\\
        \hline 
        \rowcolor{Gray}
         \textbf{Database} & \textbf{Search}  \tabularnewline
        \hline 
        IEEE & (``Industry 4.0'' OR ``Additive Manufacturing'' OR ``Augmented Reality'' OR ``Digital Twin'' OR ``Autonomous Robots'' OR ``Robotics'' OR IoT OR ``Internet of Things'' OR ``Big Data'' OR ``Cloud Computing'' OR ``Cybersecurity'' OR ``Horizontal Integration'' OR ``Vertical Integration'') AND (``Process Planning'' OR ``Quality Control'' OR ``Predictive Maintenance'' OR ``Logistics'' OR ``Robotics'' OR ``Learning Systems'' OR ``Process Control'' OR ``Process Optimization'') AND ( ``Responsible AI'' OR ``Trustworthy AI'' OR ``Transparency'' OR ``Explainability'' OR ``Traceability'' OR ``Human Agency'' OR ``Human Oversight'' OR ``Privacy'' OR ``Data Governance'' OR ``Accountability'' OR ``Fairness'' OR ``well-being'' OR ``Technical Robustness'' OR ``Safety'') AND ( ``AI'' OR ``Artificial Intelligence'')   \tabularnewline\hline
        
         Scopus & (Industry 4.0'' OR ``Additive Manufacturing'' OR ``Augmented Reality'' OR ``Digital Twin'' OR ``Autonomous Robots'' OR ``Robotics'' OR ``IoT OR Internet of Things'' OR ``Big Data'' OR ``Bigdata'' OR ``Cloud Computing'' OR ``Cybersecurity'' OR ``Horizontal Integration'' OR ``Vertical Integration'') AND (``Process Planning'' OR ``Quality Control'' OR ``Predictive Maintenance'' OR ``Logistics'' OR ``Robotics'' OR ``Learning Systems'' OR ``Process Control'' OR ``Process Optimization'') AND (``Responsible AI'' OR ``Trustworthy AI'' OR ``Transparency'' OR ``Human Agency and Oversight'' OR ``Privacy and Data Governance'' OR ``Accountability'' OR ``Fairness'' OR ``Well-being'' OR ``Technical Robustness and Safety'') AND (``AI'' OR ``Artificial Intelligence'') \tabularnewline
                        \hline

        ACM & (``Industry 5.0'' AND ``AI'') OR ((``Industry 4.0'' AND (``Industry'' OR ``Manufacturing'' OR ``Enterprise'')) AND ((``Ethics'' AND ``AI'') OR (``Sustainability'' AND ``AI'') OR ``Responsible AI'' OR ``Trustworthy AI'')) \tabularnewline
                        \hline

        GS1 & ``Trustworthy AI'' OR ``Responsible AI''  \tabularnewline
                \hline
                
        GS2 & (``Manufacturing'' OR ``Industry'')(``AI'' OR ``Artificial Intelligence'')(``responsible AI'' OR ``Trustworthy AI'')(``Industry 4.0'' OR ``Additive Manufacturing'' OR ``Augmented Reality'' OR ``Autonomous Robots'' OR ``IoT'' OR ``Big Data'' OR ``Cybersecurity'')  \tabularnewline
         
        \tabularnewline\hline
        
        GS3 & (``Industry'' OR ``Manufacturing'')(``Additive Manufacturing'')(``AI'' OR ``Artificial Intelligence'')  \tabularnewline\hline

        GS4 & (``Industry'' OR ``Manufacturing'')(``Simulation'' OR ``Modelling'')(``AI'' OR ``Artificial Intelligence'') \tabularnewline
        
        GS4\_2 & (``Industry'' OR ``Manufacturing'')(``Digital Twin'')(``AI'' OR ``Artificial Intelligence'') \tabularnewline\hline
                
        GS5 & (``Industry'' OR ``Manufacturing'')(``Augmented Reality'')(``AI'' OR ``Artificial Intelligence'')  \tabularnewline\hline
        
        GS6 & (``Industry'' OR ``Manufacturing'')(``Robot'' OR ``Robotics'')(``AI'' OR ``Artificial Intelligence'') \tabularnewline\hline

        GS7 & (``Industry'' OR ``Manufacturing'')(``IoT'' OR ``Internet of Things'')(``AI'' OR ``Artificial Intelligence'') \tabularnewline\hline

        GS8 & (``Industry'' OR ``Manufacturing'')(``Big Data'' OR ``Bigdata'')(``AI'' OR ``Artificial Intelligence'') \tabularnewline\hline        

        GS9 & (``Industry'' OR ``Manufacturing'')(``Cloud Computing'')(``AI'' OR ``Artificial Intelligence'') \tabularnewline\hline

        GS10 & (``Industry'' OR ``Manufacturing'')(Cybersecurity'')(``AI'' OR ``Artificial Intelligence'') \tabularnewline\hline

        GS11 & (``Industry'' OR ``Manufacturing'')(``Vertical Integration'' OR ``Horizontal Integration'')(``AI'' OR ``Artificial Intelligence'') \tabularnewline
        \hline
        
\end{longtable}
\end{small}

\subsection{Bibliometric Analysis}

Table \ref{table:search_results} presents the findings before the screening phase, structured on publications per year.

The last three columns of the table show: the total number of publications for the analyzed period, labelled as \textit{Total}; the correlation of $GS1$ -- as per Table \ref{table:search} -- with each query individually, labelled as $Corr_{GS1}$; and finally, the last column labelled as \textit{\%}, shows to the reader the relative percentage of queries that contain terms related to Industry 4.0 enablers -- i.e. $GS3$, $GS4\_2$, and $GS5$-$GS11$. 

\begin{table}[ht]
\caption{Searches in IEEE, Scopus, ACM and Google scholar databases between the years 2012 and 2022}\label{table:search_results}
    \begin{centering}
     \resizebox{\columnwidth}{!}{
        \begin{tabular}{|p{1.2cm}|p{1cm}|p{1cm}|p{1cm}|p{1cm}|p{1cm}|p{1cm}|p{1cm}|p{1cm}|p{1cm}|p{1cm}|p{1cm}|p{1.1cm}|p{1.1cm}|p{1.1cm}|}
        \hline 
         Database & 2012 & 2013 & 2014 & 2015 & 2016 & 2017 & 2018 & 2019 & 2020 & 2021 & 2022 & Total & $Corr_{GS1}$ & \%   \tabularnewline\hline 
        IEEE & 29 & 26 & 19 & 30 & 31 & 51 & 137 & 202 & 266 & 439 & 305 & 1535 & 0.962 & -  \tabularnewline \hline      
        Scopus & 5 & 5 & 12 & 15 & 51 & 63 & 94 & 97 & 127 & 155 & 160 & 788 & 0.862 & - \tabularnewline \hline     
        ACM & 0 & 0 & 0 & 0 & 3 & 5 & 13 & 13 & 21 & 24 & 18 & 97 & 0.873 & - \tabularnewline \hline
        GS1 & 8 & 5 & 10 & 10 & 17 & 21 & 106 & 875 & 2320 & 4500 & 3430 & 11302 & 1.000 & - \tabularnewline \hline      
        GS2 & 1 & 1 & 1 & 2 & 9 & 6 & 46 & 347 & 859 & 1640 & 1750 & 4662 & 0.981 & - \tabularnewline \hline       
        GS3 & 177 &	326 & 553 &	887 & 1410 & 2150 & 3610 & 5140 & 4360 & 10900 & 9900 & 42413 & 0.959 & 0.048 \tabularnewline	\hline       
        GS4 & 73200 & 82000 & 87300 & 97600 & 101000 & 103000 & 94900 & 96700 & 99700 & 74900 & 122000 & 1032300 & 0.121 & - \tabularnewline    
        GS4\_2 & 28 & 23 & 38 & 26 & 47 & 239 & 807 & 1870 & 3710 & 6970 & 7230 & 20988 & 0.965 & 0.024 \tabularnewline \hline       
        GS5 & 898 & 1110 & 1240 & 1580 & 1990 & 3470 & 5450 & 8230 & 11200 & 15000 & 12200 & 62368 & 0.943 & 0.070 \tabularnewline \hline	
        GS6 & 11800 & 15500 & 16900 & 17700 & 19400 & 23600 & 32800 & 46100 & 48600 & 44400 & 42800 & 319600 & 0.766 & 0.361 \tabularnewline \hline       
        GS7 & 940 & 1330 & 1950 & 3380 & 5590 & 11200 & 18100 & 23900 & 31900 & 42200 & 37900 & 178390 & 0.923 & 0.201 \tabularnewline	\hline       
        GS8 & 329 & 817 & 1640 & 2750 & 4140 & 7190 & 12500 & 16200 & 17000 & 17500 & 16300 & 96366& 0.777 & 0.109 \tabularnewline \hline	       
        GS9 & 1820 & 2310 & 2830 & 3360 & 4150 & 6180 & 9860 & 14200 & 17100 & 17200 & 14000 & 93010 & 0.829 & 0.105 \tabularnewline	\hline      
        GS10 & 754 & 774 & 925 & 1240 & 2240 & 3320 & 4760 & 7060 & 9470 & 13700 & 8850 & 53093 & 0.929 & 0.060 \tabularnewline	\hline      
        GS11 & 1040 & 1010 & 1040 & 1090 & 1190 & 1410 & 1880 & 2430 & 2780 & 3310 & 2550 & 19730 & 0.896 & 0.022 \tabularnewline\hline
        \end{tabular}
        }
    \end{centering}
\end{table}

The initial five rows after the header of Table \ref{table:search_results} display the queries conducted across IEEE, ACM, Scopus, and Google Scholar databases ($GS1$, and $GS2$). These queries adhere to the search approach outlined in the research methodology. It is evident that the number of publications has increased since 2018, with many queries demonstrating a significant correlation to ethical-driven work, independently to the industrial sector or specific Trustworthy AI concepts. 

The following rows in the table assess the implications for the industrial sector, targeting the pillars associated with specific concepts of Industry 4.0 (i.e., $GS3$ to $GS11$). The queries were performed with and without the inclusion of \textit{responsible AI} or \textit{Trustworthy AI} keywords related to ethical concepts. Analyzing the incorporation of ethical concepts across Industry 4.0 technological enablers, it can be observed that the overall percentage of publications incorporating these concepts was over 0.90\% of the total analyzed publications. However, when considering publications after 2018, this percentage increased to up to 3.61\%. The term \textit{Industry 4.0} and its corresponding pillars were introduced in 2011 and gained popularity after 2016 \cite{philbeck_fourth_2018}. The analysis reveals that all pillars, except for \textit{Simulation}, experienced similar periodic increases in attention from the scientific community.

As previously described, the broadness and generality of the term \textit{Industry 4.0} contribute to its disconnection from other keywords. Similarly, while \textit{Robotics and Automation} have been closely associated with Industry 4.0, their trend in the general industrial sector ($GS6$) shows a different pattern compared to other Industry 4.0 technological enablers, with a moderate correlation coefficient of 0.766. Notably, the development of the first digital and programmable robot dates back to 1954, and its incorporation across various domains has steadily increased since then.

Furthermore, based on these observations, an estimation of the relative numbers and their impact on the field of Industry 4.0 can be studied. The analysis indicates that \textit{Additive Manufacturing}, \textit{Digital Twins}, and \textit{Horizontal and Vertical Integration} are the topics least covered in the literature. It is important to consider that \textit{Additive Manufacturing} may be referred to by different names depending on the specific technology employed. Hence, it could be underrepresented unless a comprehensive search is conducted for each relevant topic. On the other hand, the concept of \textit{Digital Twins} is gaining increasing interest in the industrial sector, with recent considerations being made regarding its connection with AI \cite{castane_assistant_2022}.

Simulation and Digital Twins have some similarities in their use of virtual models and data-driven analysis, but they differ in their focus and purpose. Simulations have a broader scope and are utilized to understand and predict system behaviour. On the other hand, Digital Twins are specific to individual assets or processes and offer real-time monitoring and optimization capabilities. Although Simulations are widely known, the emergence of Digital Twins has gained significant attention due to their ability to leverage data, analytics, and connectivity for optimizing industrial processes, enhancing efficiency, and enabling human-machine collaboration. These concepts, along with their impact on human-machine collaboration, life-cycle management, remote monitoring and control, optimization of industrial systems, and improvements in efficiency and productivity, have captured widespread interest, particularly within the context of Industry 5.0 and, thus, the growing capture of interest as observed in GS4\_2.

Finally, \textit{Horizontal and Vertical Integration} may have limited considerations in terms of AI. These topics are more closely associated with communication methods and the management of systems of systems (e.g., Edge Computing, Cloud Computing), and their direct linkage to AI could be relatively low.

\subsection{Screening papers}
\begin{figure*}[htbp!]
    \centering
    \includegraphics[width=137mm]{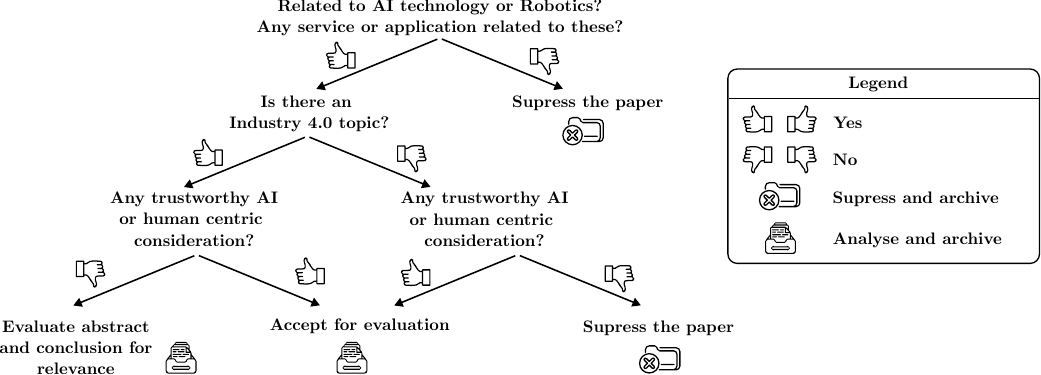}
    \caption{Screening process followed to select the papers based on the keywords}
    \label{fig:screening}
\end{figure*}

A systematic analysis was conducted on the collected dataset with a primary focus on assessing the relevance of the titles to the key facets identified previously. In cases where uncertainty arose, a more detailed analysis was performed based on the information provided in the abstracts. This evaluation process followed a structured pipeline approach, as illustrated in Figure \ref{fig:screening}. The main objective of this rigorous screening was to ensure that each manuscript included at least one Trustworthy AI requirement along with an additional topic related to either AI in industry or Industry 4.0 technological enablers. Manuscripts solely addressing AI ethics were excluded from consideration. Furthermore, manuscripts explicitly referencing Industry 5.0 were also excluded from the analysis to maintain a focused examination of the relevant topics of interest.

\subsection{Keywording using abstracts}

The predefined set of keywords served as the basis for conducting the mapping and analysis processes in the study. However, during the screening process, and together with the keywords associated with Trustworthy AI requirements, three distinct technologies/methods emerged that were deemed relevant to the investigation. These technologies/methods include \textit{Blockchain}, \textit{Federated Learning}, \textit{MLOps}, and Big Data Stream Processing. These will be described and thoroughly discussed in the forthcoming results and discussion section of the study.

\subsection{Data extraction and Mapping process}\label{results and findings}

This section presents the findings of the systematic literature review following the rigorous screening process. The material provides comprehensive details on the analysis sets, including the sources utilized for extracting research questions and findings. Additionally, the results were enhanced with bibliometric analyses to improve the understanding of the queried information.

In each Research Question, a succinct discussion has been incorporated to establish a direct correspondence with the information presented in the figures. This preliminary discussion serves as a precursor to the subsequent section, where a detailed analysis of the findings, future directions, and identified research gaps is presented.

The outcomes are depicted in grid charts in the upcoming sections. The values used represent the counts of a specific topic. These numbers on the intersections are the individual manuscripts where the topics intersected were detected. The first row and last column represent the cumulative count where all topics are aggregated in the specific facet. 

Furthermore, two additional categories were added to the mapping on each facet: general overview, and technique specific. General overview refers to manuscripts with a generalization of the aspects of AI and, thus even when they can be relevant as a research topic, do not provide insights on topics related to AI. There were also considered in this category those focused on Trustworthy requirements or applications of AI without a thorough applicability within the Industry 4.0 enablers. Technique-specific refers to manuscripts covering the application of a concrete AI technique (e.g. Random Forest, 2D Convolutional Neural Network) or discussion of the implementation of different AI life-cycles (e.g. Training) within the industrial sector but not in a specific industrial application (e.g. Process planning).

Table \ref{TABLE:EXTRA}, summarizes the number of mentions within each of the evaluated manuscripts with respect to the total. Given the size limitations of the paper, the topics have been consolidated within nine different options for each one of the columns as described in the caption. 

\begin{table}[h!]
\caption{Total Numbers of References in Each Topic}
\label{TABLE:EXTRA}
\resizebox{\textwidth}{!}{%
\begin{tabular}{|>{\centering\arraybackslash}m{4cm}|>{\centering\arraybackslash}m{2cm}|>{\centering\arraybackslash}m{2cm}|>{\centering\arraybackslash}m{2cm}|>{\centering\arraybackslash}m{2cm}|>{\centering\arraybackslash}m{2cm}|>{\centering\arraybackslash}m{2cm}|>{\centering\arraybackslash}m{2cm}|>{\centering\arraybackslash}m{2cm}|>{\centering\arraybackslash}m{2cm}|}
\hline
\rowcolor{Gray}
\textbf{Topic} & \textbf{Research Type} & \textbf{Trustworthy AI} & \textbf{Industry 4.0} & \textbf{AI Application} \\
\hline
Topic 1 & 75 & 75 & 36 & 76 \\
\hline
Topic 2 & 137 & 126 & 1 & 2 \\
\hline
Topic 3 & 10 & 108 & 12 & 7 \\
\hline
Topic 4 & 51 & 104 & 25 & 12 \\
\hline
Topic 5 & 29 & 18 & 126 & 15 \\
\hline
Topic 6 & 6 & 28 & 117 & 115 \\
\hline
Topic 7 & - & 23 & 81 & 36 \\
\hline
Topic 8 & - & - & 24 & 100 \\
\hline
Topic 9 & - & - & 48 & 14 \\
\hline
\end{tabular}%
}

\smallskip
{\footnotesize \textbf{Topics from top to bottom} \textit{Research Type:} Solution, Validation, Evaluation, Phylosophical, Review, Experience. \textit{Trustworthy AI:} Human Agency \& Oversight, Robustness \& Safety, Privacy \& Data Governance, Transparency, Diversity, Non-Disc. \& Fairness, Societal \& Env. Wellbeing, Accountability. \textit{Industry 4.0:} General Industry, Additive Manufacturing, Augmented Reality, Simulation, Autonomous System, IOT \& Big Data, Cloud Computing, Cybersecurity, Horizontal \& Vertical Integration. \textit{AI Application:} Abstract AI, Process Planning, Quality Control, Predictive Maintenance, Logistics, Robotics, Assistance \& Learning Systems, Concrete AI, Process Control \& Optimisation }

\end{table}

\subsection*{RQ1 - What are the key Trustworthy AI requirements prioritized in Industry 4.0 technological enablers and their AI applications, and how can they be further promoted?}

\begin{figure*}[htbp!]
    \centering
    \includegraphics[width=90mm]{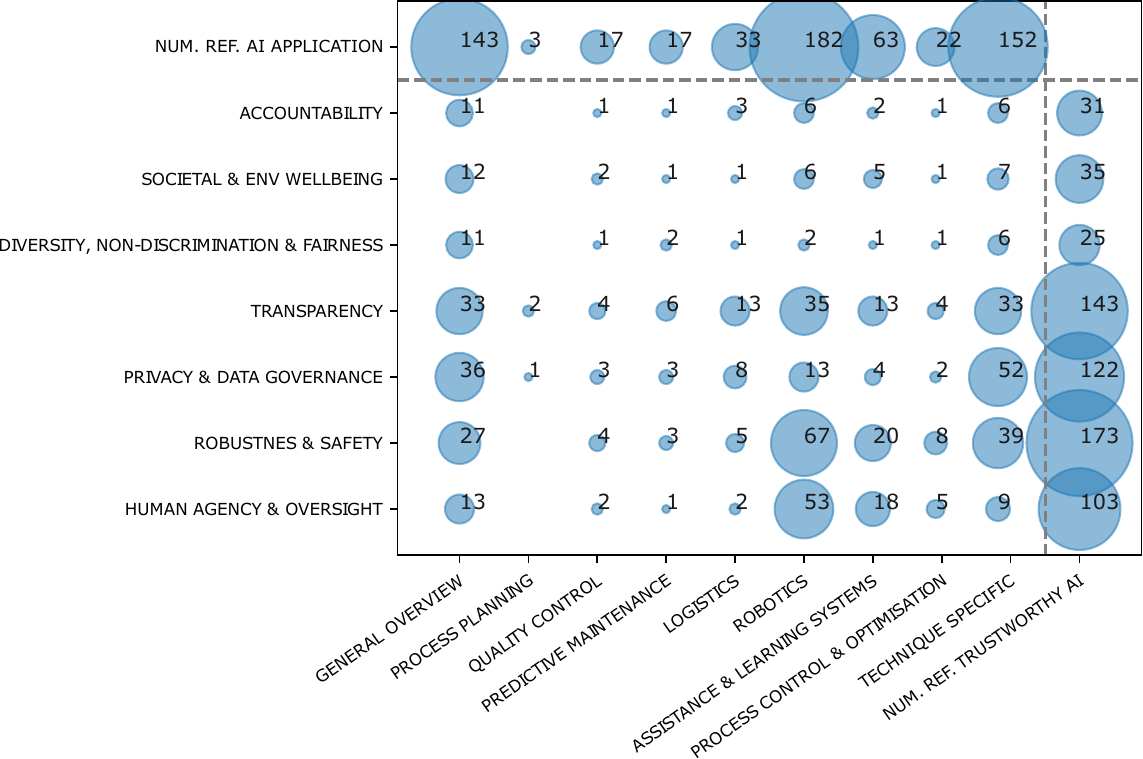}
    \caption{Trends in AI application vs Trustworthy AI}
    \label{fig:RQ1_2}
\end{figure*}

The Trustworthy AI topics that have garnered significant attention are those that exhibit strong associations with technical components, as depicted in Figure \ref{fig:RQ1_2}. 
Among these concepts, Robustness and Safety have received the highest interest (27.37\%), followed by Transparency (22.63\%), and Privacy and Data Governance (19.3\%). Notably, Human Agency and Oversight occupy the fourth position (16.3\%). This trend can be attributed to the widespread integration of human-in-the-loop, human-on-the-loop, and human-in-control considerations in Autonomous Systems. In Robotics, a substantial proportion of manuscripts focus on human interaction and emphasize robustness and safety considerations (36\% on 67 identified manuscripts across 182 mentions on Robotics). Noteworthy findings in this context can be observed in the domain of autonomous vehicles \cite{khan_design_2019,muzahid_optimal_2021,wu_learning_2019}. Furthermore, the transportation industry has displayed a growing interest in other Trustworthy AI topics, such as transparency \cite{schenk_creating_2020,jaworek-korjakowska_safeso_2021}. However, as expected in industries with high utilization of Autonomous Systems, the primary focus revolves around Technical Robustness, Safety \cite{beetz_robotic_2015,el_maadi_suspicious_2013,clark_malicious_2018}, and topics related to human-centric approaches, human-machine collaboration, and decision-making \cite{keshavdas_functional_2012, baiden_human-robot-interaction_2013, rozo_learning_2014}.

Another area of high interest in the literature within the context of Industry 4.0 is the combination of IoT, Big Data, and Cloud Computing. The integration of Big Data analytics and AI in the industrial sector is a common topic of discussion \cite{roh_survey_2021,pencheva_big_2020}. However, these technologies also raise concerns regarding privacy, particularly when personal data is collected and utilized. The use of sensitive data, such as process data, corporate data, healthcare data, and financial data, introduces additional security and privacy challenges. Furthermore, various communication channels, including social media platforms and Edge/Cloud services, can pose potential privacy threats. This consideration becomes even more significant and complex within the Industry 4.0 framework, which fosters the development of AI-driven intelligent applications within a network interconnected and real-time-based system.

On the other hand, the Trustworthy requirements that have received relatively less attention in relation to Industry 4.0 technological enablers are Diversity, Non-discrimination, and Fairness (3.95\%), Accountability (4.9\%), and Societal and Environmental Well-being (5.56\%). 

Within diversity, non-discrimination, and Fairness requirement, the topic of Fairness is the main focus of research \cite{liu2019adversaries,zhao2019machine,li2020blockchain,ur2021trustfed,kenfack2021fairness}. The relevance of diversity and non-discrimination within the industrial sector may be relatively lower compared to other topics. However, as these aspects are further explored and integrated into other domains -- e.g. healthcare, human labour, and education --, their consideration and implementation within Industry 4.0 can be expected to be promoted.

Accountability involves establishing mechanisms to ensure responsibility and accountability for AI systems and their outcomes. As defined in the Trustworthy AI guidelines, accountability encompasses topics such as auditability, minimization and reporting of negative impacts, redress, and specification of trade-offs between different Trustworthy AI requirements. Applied to AI Applications, only the 4.9\% is identified split in between general overview (7.7\%), assistance and learning systems (3.23\%), and technique specific (3.95\%). The accountability field is in between a scientific and management area and therefore requires of heterogeneous skills to have a broad understanding to innovate on it. However, this is a key requirement for future legislation. Some considerations are related to AI standardisation -- existing and future.

The management process plays a crucial role in effectively handling accountability and can be facilitated through the use of management tools or components to aid in the implementation, development, and utilization of AI systems. Moreover, considering the risks associated with AI systems, the use of risk management tools can further enhance the development of new approaches. For instance, various frameworks and standards are currently being developed in this regard \cite{vyhmeister2023risk, vyhmeister2023responsible}. By contrast, recent efforts were made to establish standards that promote accountability and cover other Trustworthy AI topics. For example, the IEEE 7001-2021 standard for Transparency of Autonomous Systems \cite{ieee_ieee_2021}. This standard is applicable to Autonomous Systems, including automated driving systems and assisted living robots, and addresses issues including recommender systems and chatbots used for medical diagnosis. It specifically focuses on the potential to cause physical, psychological, societal, economic, environmental, or reputation harm.

Finally, regarding societal and environmental well-being (5.56\%), as mentioned in the IEEE standard for transparency \cite{ieee_ieee_2021}, the impact on society and the environment can be associated with indirect access to AI data or processes, such as unauthorized data access. Therefore, implementing robust privacy and data management processes can contribute to societal and environmental well-being. As different regulatory conditions are developed and implemented for AI assets, the potential harm resulting from specific risks can be minimized. Consequently, as policies are established, the industry will regulate itself, reducing these risks.

Overall, when examining the trends of Trustworthy AI across different AI applications of the Trustworthy requirements that have received the most attention are those associated with technical components -- Transparency, Privacy \& Data Governance, and Robustness \& Safety. 
Additionally, Human agency and oversight also exhibit a relatively high level of coverage, primarily contributed by Robotics AI assets that involve considerations of human-in-the-loop, human-on-the-loop, and human-in-command. Given the significance of IoT, Autonomous Systems, and Cloud Computing, it is not surprising that the most relevant AI applications are closely linked to these drivers and, at the same time, demonstrate relevance to Trustworthy AI considerations, such as Robotics, Machine Learning, and assistance and learning systems. 
Conversely, the AI applications that exhibit the lowest relevance, such as process planning, quality control, logistics, and predictive maintenance are already well-established topics in the industrial setting. However, the results clearly indicate that Trustworthy AI has not yet permeated these areas, highlighting the need to focus attention on understanding the impact of Trustworthy AI requirements on each of these specific AI applications.

\subsection*{RQ2 - What are the current trends in the relationship between Industry 4.0 enablers and AI applications?}

Next Figure \ref{fig:RQ2} provides a classification of AI applications in industrial domains based on the Industry 4.0 Technology Enablers. The selection process for the papers ensured that manuscripts addressing Trustworthy AI requirements, even indirectly, were included.

\begin{figure*}[htbp!]
    \centering
    \includegraphics[width=90mm]{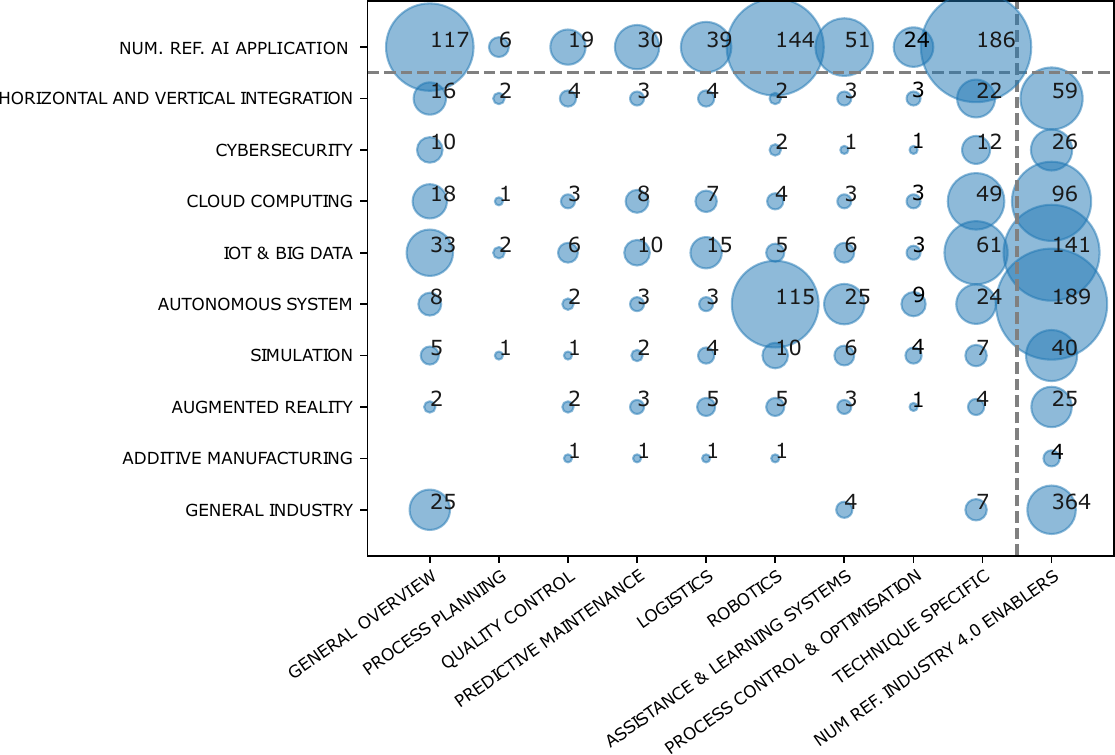}
    \caption{Trends in AI application vs industry 4.0 technological enabler}
    \label{fig:RQ2}
\end{figure*}

Without considering general overview papers and technique specific (19\% and 30.2\% respectively), Robotics (23.28\%) and assistance and learning systems (8.28\%) are the main areas that attracted significant research interest in AI applications. It is noticeable the amount of work that can be found in general descriptions and road mapping as well as in techniques that are not into any concrete industry enabler. Furthermore, most of the research on Robotics applications focuses on Autonomous Systems, with minimal contributions to other Industry 4.0 Technology Enablers.

The adoption of Robotics has substantially increased in industrial environments with applications ranging from welding and disassembly to pick and place and transportation. Cooperative robots, known as Cobots, have emerged as a new trend in Autonomous Systems, having physical interactions with humans \cite{sivan_cobots_2021}. Safety considerations play a crucial role in human-robot interactions, and they are integral to Trustworthy requirements. As depicted in Figure \ref{fig:RQ1_2}, major efforts in Robotics are around robustness \& safety is addressed with a 36.8\%, Human Agency \& Oversight with a 29.1\%, and Transparency a 19.2\%. Ensuring safety involves measures to prevent direct physical contact between humans and robots, considerations of decision responsibility, human agency and oversight in interactive systems, and transparency to improve worker adoption. Moreover, several studies have focused on novel techniques for ensuring human safety during interactions with robots\cite{sidner_where_2004,costanzo_multimodal_2022,eder_towards_2014}. Often these are based on predicting the engagement state of humans and robots, improvements in these models, including predictions of human intentions, contribute to enhancing security~\cite{abdelrahman_multimodal_2022}.

In contrast to Robotics, technique-specific papers are a big representative in the classification, published across different Industry 4.0 drivers, particularly remarkable in information distribution such as IoT \& Big Data (37.8 \%) and Edge-Cloud Computing (26.3 \%).

AI adoption aims to improve the robustness and quality of industrial processes. However, incorporating AI and ML applications presents the challenge of user acceptance. Consequently, efforts to unveil the black-box models have gained significant attention in the sector \cite{ogrezeanu_privacy-preserving_2022}. This is illustrated in Figure \ref{fig:RQ1_2} on the ML techniques encompassed within technique-specific with respect to the fields of Robustness (22.5\%), Transparency (23.1\%) and Privacy and Data Governance (42,6\%), in the industrial sector and about 27.3\%, 22.6\%, and 19.3\% of the overall interest in Trustworthy requirements. Thus, the importance of ML within technique-specific in different AI applications, considering Trustworthy considerations, is evident.
In addition, new trends in Reinforcement learning processes have shown promise in enhancing Safety and Human Agency and Oversight in complex environments \cite{li_safe_2021,muzahid_optimal_2021,bastani_safe_2021,iucci_explainable_2021}.

Among the AI applications, Augmented Reality (2.64\%) and Additive Manufacturing (0.42\%) exhibit lower importance. Augmented Reality is primarily applied in human-robot collaboration tasks, maintenance, and assembly \cite{reljic_augmented_2021}, with minor contributions to manufacturing, training, and logistics in Industry 4.0. However, as pointed out by \cite{reljic_augmented_2021}, Industry 4.0 is currently focused on resolving technical challenges, and aspects such as ergonomics, technological complexity, and application usability is able to incorporate Augmented Reality.

In the case of Additive Manufacturing, the limited number of findings could be attributed to different nomenclatures. Nonetheless, the application of Additive Manufacturing with Trustworthy AI considerations requires overcoming various challenges, including data complexity, heterogeneity, explainability, limited data availability, cyber-physical security and privacy, human agency and oversight, and the integration of Augmented Reality for quality processing tasks such as interaction with 3D geometry for quality inspection \cite{liu_when_2022}. These challenges represent a convergence of previously mentioned limitations within a specific AI application.



\subsection*{RQ3 - What innovations can facilitate the incorporation of Trustworthy AI guidelines within Industry 4.0 towards Industry 5.0?} 
\label{RQ3}

To assess the extent to which Trustworthy requirements have been integrated into AI applications and Industry 4.0 Technology Enablers, an analysis of the readiness of these applications was conducted using the framework proposed by Petersen et al. \cite{petersen2008systematic} as shown in Table \ref{table:research_facet}. To these categories, a new facet -- Opinion Paper -- is proposed to encompass the manuscripts with a higher level of abstraction than Evaluation, such as the surveys.

The quantitative measure used for the incorporation of Trustworthy AI requirements within Industry 4.0 is the Technology Readiness Level (TRL) grade, used to evaluate the manuscripts. The TRL scale ranges from TRL 0 with the first principles, to TRL 9 where the system is implemented or deployed in real environments. The intermediate levels represent different stages of development and validation according to \cite{lavin_technology_2021}. More specifically, the TRL scale is used as follows: solution proposal (TRL 1 to 2), validation (TRL 3 to 4), and evaluation (greater than 5). On the case of philosophical paper, opinion paper and experience paper, the TRL scale is not applicable as described in the Table. It is important to note that, based on our analysis, no work with a TRL higher than 6 was identified in the literature.

Furthermore, in the context of ML components, the TRL was explored in a study by Lavin et al. \cite{lavin_technology_2021}. This study used TRL to assess the readiness of ML algorithms and their associated data management processes for implementation in specific domains.

\begin{table}[h!]
\caption{Research Type Facet based on \cite{petersen2008systematic}}
\label{table:research_facet}
    \begin{centering}
     \centering{}
    \small\addtolength{\tabcolsep}{0pt}
        \begin{tabular}{|V{4.5cm}|V{8cm}|V{4.5cm}|}
        \hline 
        \rowcolor{Gray} 
         \textbf{Category} & \textbf{Description} & \textbf{TRL}  \tabularnewline
        \hline 
        Solution Proposal & An innovative or significantly extended solution is presented to address a specific problem. The solution not only highlights its uniqueness but also provides a compelling rationale or illustrative examples to demonstrate its potential benefits. & 1 and 2 \tabularnewline \hline 
        Validation & Novel techniques, methods, and algorithms are proposed and validated through theoretical analysis, without being implemented in practical use cases. & 3 and 4 \tabularnewline \hline 
        Evaluation & The implemented techniques are applied in real-world scenarios, accompanied by a thorough evaluation process. & $>=$ 5 \tabularnewline \hline 
        Philosophical Paper & Propose a fresh perspective on existing concepts by organizing the field using a taxonomy or conceptual framework.& N.A. \tabularnewline \hline 
        Opinion / Review Paper & Present an individual viewpoint concerning a technique, method, or algorithm, highlighting the potential for enhancements or modifications. Avoid relying on prior research methodologies or related work in the discussion. & N.A. \tabularnewline \hline 
        Experience Papers & Provide a detailed account of the practical implementation, outlining the specific actions taken and their execution. It is essential for the author to possess personal experience as a prerequisite for this description.  & N.A. \tabularnewline 
        \hline
        \end{tabular}
    \par\end{centering}
\end{table}

\begin{figure*}[htbp!]
\begin{minipage}[c]{0.49\linewidth}
    \centering
    \includegraphics[width=65mm]{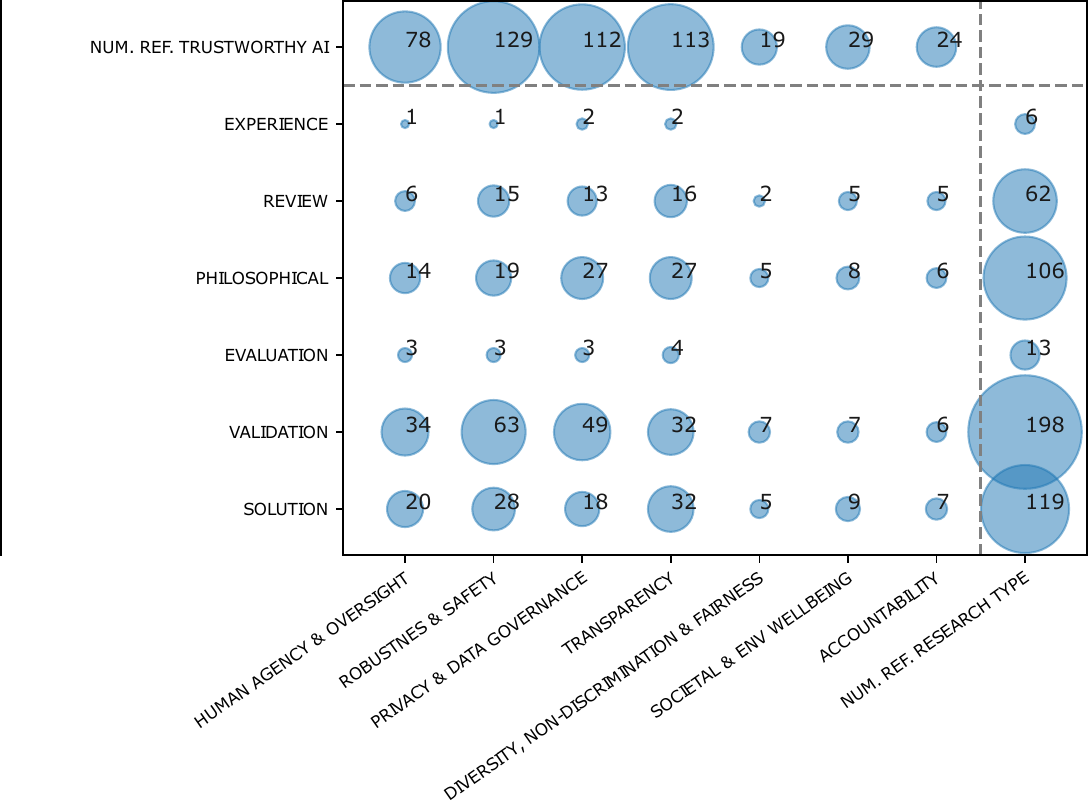}
    \caption{Facet vs Trustworthy AI Requirement}
    \label{fig:RQ3_1}
\end{minipage}
\begin{minipage}[c]{0.49\linewidth}
    \centering
    \includegraphics[width=67mm]{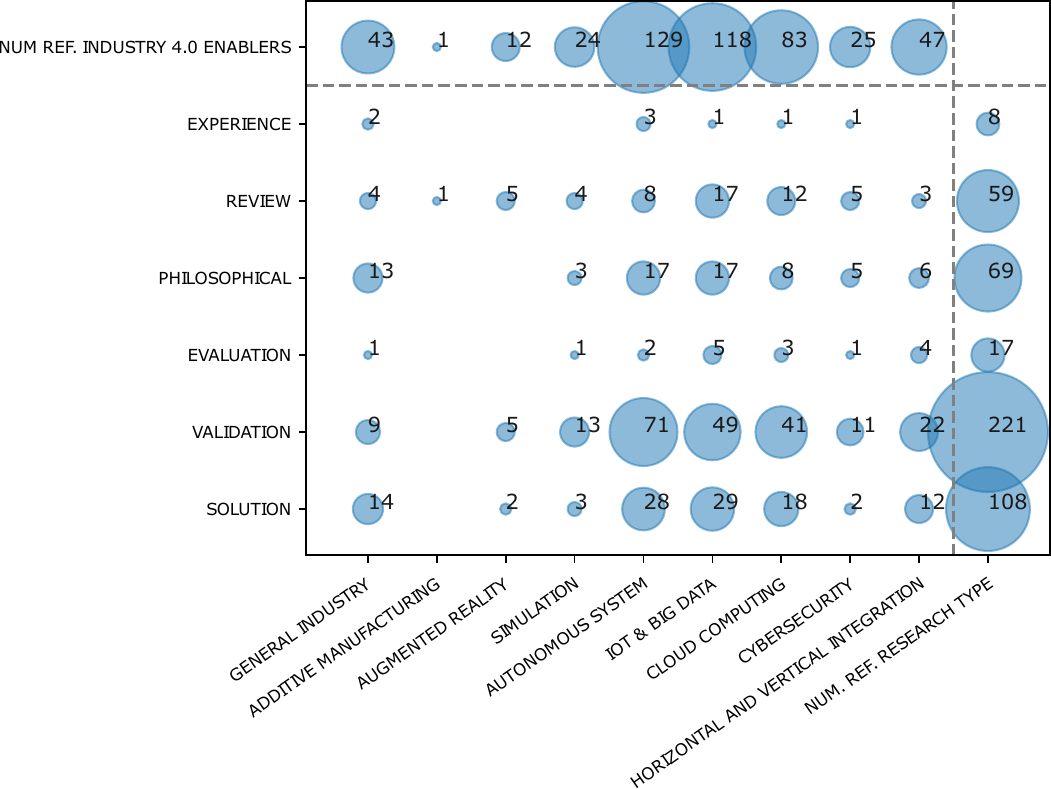}
    \caption{Facet vs Industry 4.0 technological enabler}
       \label{fig:RQ3_2}
\end{minipage}
\end{figure*}

\begin{figure*}[htbp!]
    \centering
    \includegraphics[width=66mm]{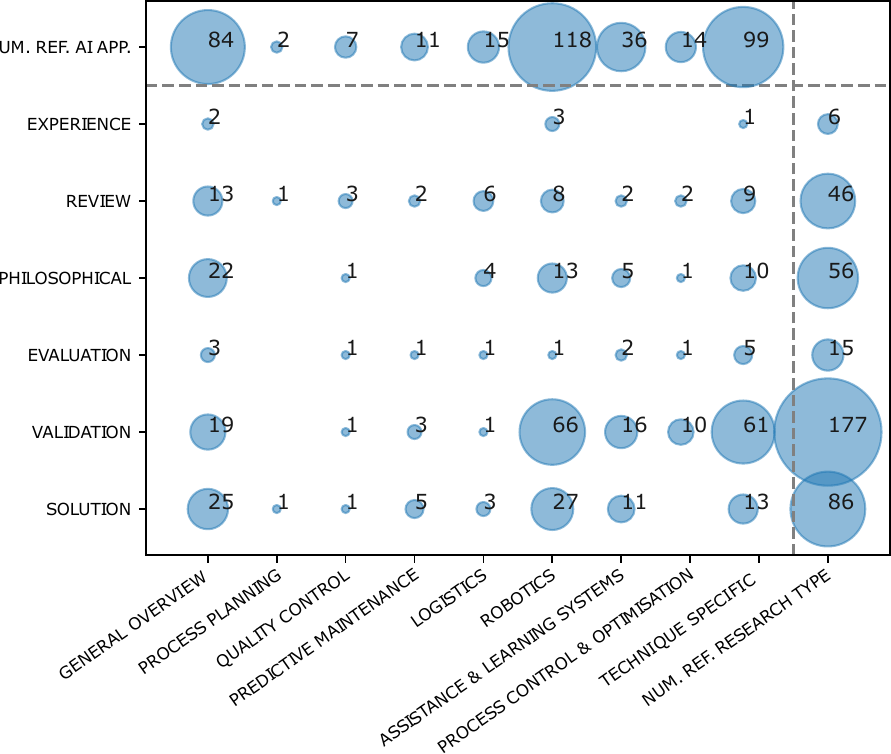}
    \caption{Facet vs Trends in AI applications}
    \label{fig:RQ3_3}
\end{figure*}

Figures \ref{fig:RQ3_1} to \ref{fig:RQ3_3} show the trends of the research type vs the Trustworthy requirements, the Industry 4.0 Technology Enablers, and the AI applications domain, respectively.

From a general perspective, the analysis of the Figures reveals a common trend in the papers based on their TRL. The majority of papers fall into the validation category, accounting for 39.3\% for Trustworthy AI, 46\% for Industry 4.0 Technology Enablers, and 45.9\% for Trends in AI applications. This is followed by the solution category with 23.6\%, 22.5\% and 22.2\% in the same order as earlier described. Finally, the evaluation category with 2.6\%, 3.5\% and 3.9\% respectively. It is evident that TRL 3 is the most prevalent, indicating that the analyzed work primarily consists of proof of concept and validation stages. Consequently, it is still in an early stage of development and not yet operational for practical implementation in companies and industries. It is also important to notice that the numbers reflect the lower adoption in industry on some of these topics and that they are not yet ready to be used in some of the real scenarios. This semi-qualitative assessment provides a general understanding of the progress towards the transition to Industry 5.0. However, a more comprehensive analysis can be conducted for each specific topic within the Trustworthy AI requirements.

Another noteworthy trend observed during the analysis is the inclusion of ethical and trust considerations in models to make them operational and linked to different stages of development. These models serve the purpose of enhancing the understanding of specific requirements and providing metrics for managing ethical risks. Some notable findings include the development of a surrogate model for explainability \cite{tan_surrogate_2022}, a robot-to-human trust model \cite{wang_computational_2022}, and a model for transparency between emerging technologies and humanitarian logistics sustainability \cite{khan_model_2022}. As these models continue to advance, a better understanding of the collaboration between AI-driven assets and human counterparts can be achieved.

\section{Discussion}
\label{discussion}

\subsection{Trends and enabler technologies to transition to Industry 5.0}

During the assessment of the manuscripts, novel trends were identified as facilitators for the transition from Industry 4.0 to Industry 5.0. These emerging trends encompass Blockchain, Federated Learning, ML operations (MLOps), and Big Data Stream Processing.

\begin{figure*}[h!]
    \centering
    \includegraphics[width=85mm]{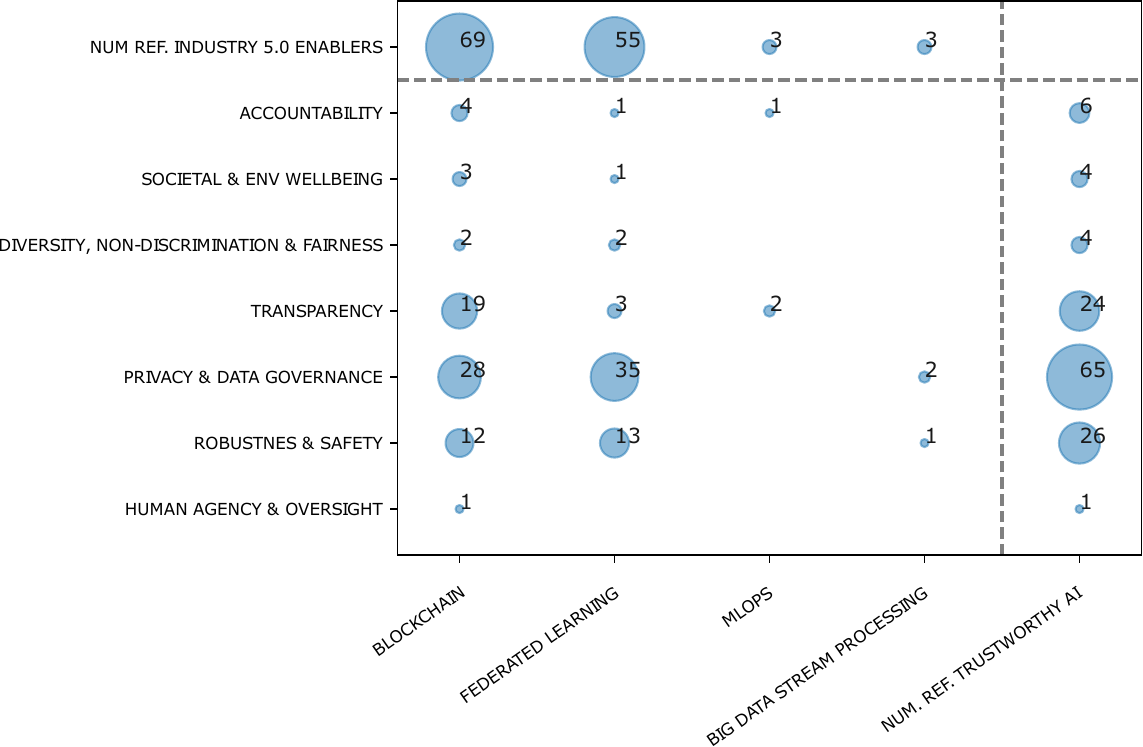}
    \caption{Emergent Trends in Trustworthy AI Requirements}
    \label{fig:RQ3_4}
\end{figure*}

\begin{figure*}[h!]
    \centering
    \includegraphics[width=85mm]{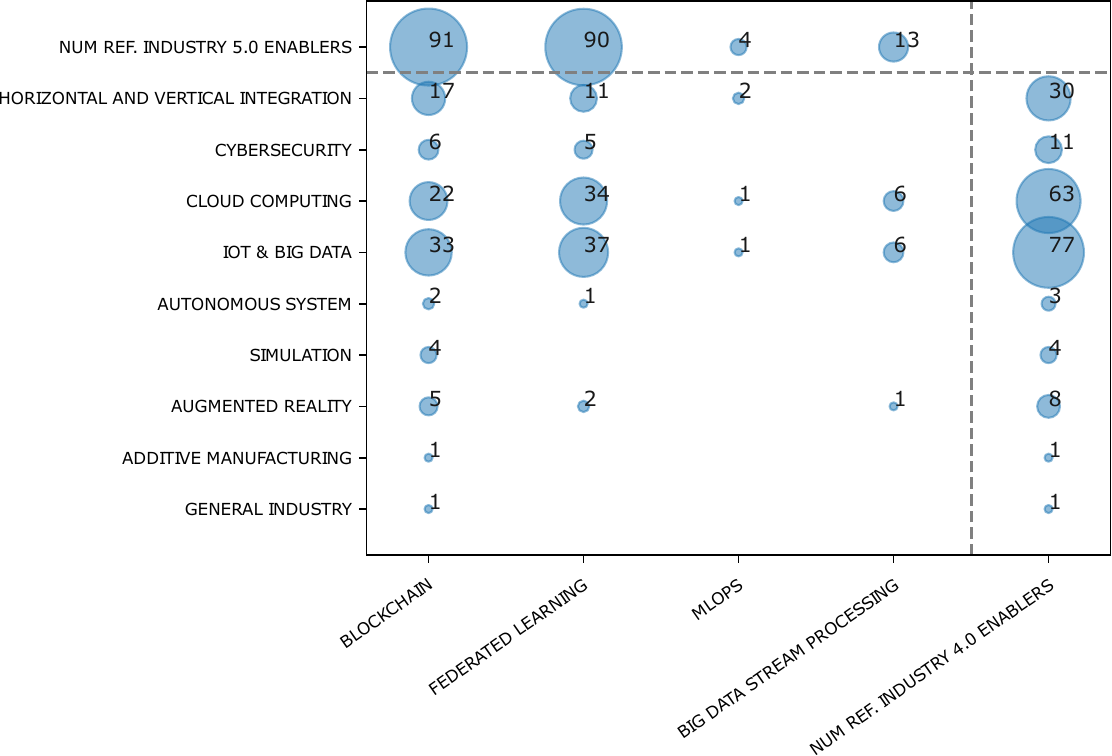}
    \caption{Emergent trends in Industry 4.0 Technology Enablers}
    \label{fig:RQ3_5}
\end{figure*}

\begin{figure*}[h!]
    \centering
    \includegraphics[width=85mm]{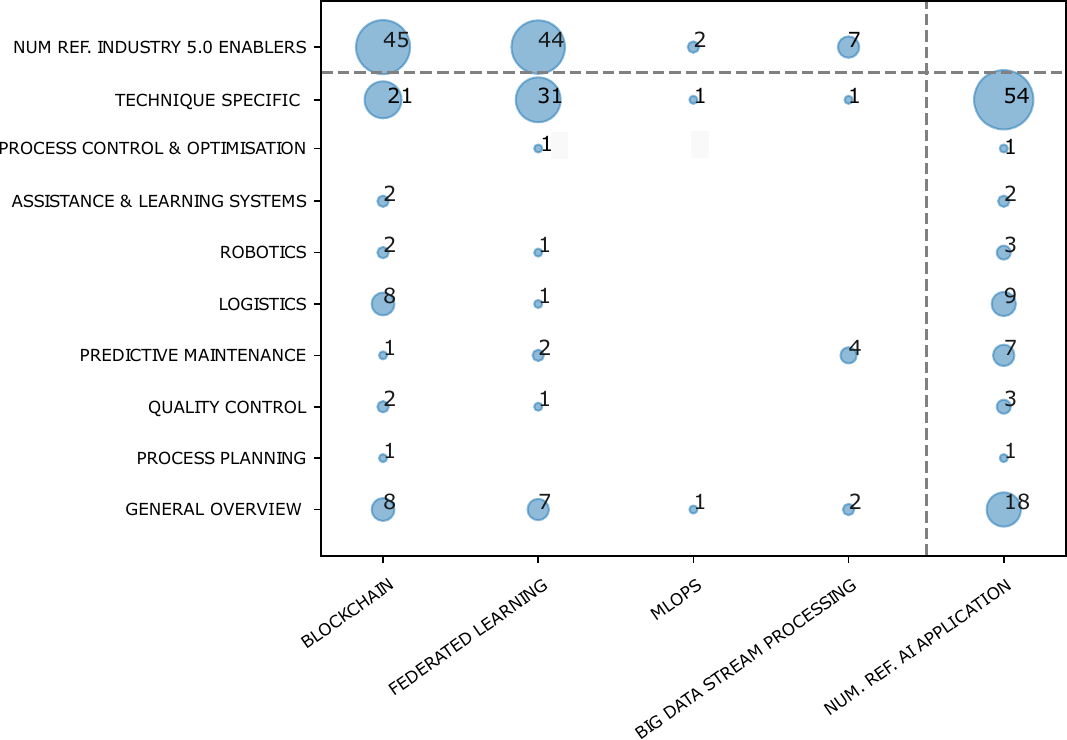}
    \caption{Emergent Trends in AI Applications in Industry 4.0}
    \label{fig:RQ3_6}
\end{figure*}

These emerging trends are shown in the following figures following the same approach as previously presented with a specific focus on comparing the four major emergent topics with the Trustworthy AI requirements in Figure \ref{fig:RQ3_4}; Industry 4.0 Technology Enablers in Figure \ref{fig:RQ3_5}; and AI applications in Figure \ref{fig:RQ3_6}. The relations will be described next in the Section.


Figure \ref{fig:RQ3_4} shows the relation of the emergent trends and its relation with Trustworthy AI requirements. Blockchain and Federated Learning have similar trends with major contributions to Privacy \& Data Governance, and Robustness \& Safety. The main difference between them is Transparency, where Blockchain defines different benefits. MLOps only had a few contributions as described in the literature but with a strong potential given the need for software standard methodologies in the industrial sector. In the same manner, Big Data Stream Processing presents only a few works however it has the potential to be a key enabler in the transition as will be explained later in the Section.
 
Figure \ref{fig:RQ3_5} depicts the mentioned methods' contribution to the Industry 4.0 Technology Enablers. Blockchain and Federated Learning describe similar contributions in which the communication and data transfer/managing systems are considered (i.e. IoT, Cloud or Edge Computing, Horizontal \& Vertical Integration, and Cybersecurity).

Figure \ref{fig:RQ3_6} presents the analysis of the contributions of AI applications within the industrial domain. Since these methods deal with approaches of accessing and managing distributed data and information for AI training, the Figure shows that the implementation of specific techniques and the discussed emergent technologies grow in interest and therefore will have a big impact on industry as drivers for the transition to Industry 5.0.

\subsubsection{Blockchain}
As highlighted by Mushtaq et al. \cite{mushtaq_implications_2019}, cryptographic techniques were widely employed to safeguard sensitive information. Originally developed for information security purposes, these techniques have begun to permeate various domains. Notably, Blockchain technology, initially devised for the Bitcoin cryptocurrency, has emerged as a decentralized transaction and data management solution \cite{mushtaq_implications_2019}. Since its conceptualization in 2008, Blockchain has garnered significant attention owing to its inherent features that ensure security, data integrity, anonymity, and reliability. The industrial sector has witnessed a substantial increase in the adoption of Blockchain technology within the framework of Industry 4.0. It offers numerous advantages, including the support for autonomous decision-making processes and real-time utilization of streaming information \cite{mushtaq_implications_2019,fraga-lamas_review_2019,madumidha_transparency_2019,muller_hidals_2019}. The integration of intelligent sensors and actuators for enhancing industrial operations is facilitated by the Industrial IoT, which leverages smart devices and real-time analytics to harness the data collected from sensors.

The analytics and communication also deal with sensitive data, which should be protected against any information security-related attacks. Thus Blockchain and other mechanisms should be used to safely and securely process information. Furthermore, the inherently decentralized nature of Blockchain technology provides significant advantages in mitigating system failure-related issues, particularly in the context of autonomous system control. By leveraging Blockchain, the state condition of deployed tools and components can be accurately measured, thereby ensuring the robustness and reliability of the system. A seamless integration of Blockchain with AI further enhances its utility. For instance, Blockchain technology ensures the integrity and security of extensive data indexes originating from diverse sources. This secure foundation enables AI systems to effectively and comprehensively learn from the data, facilitating more accurate and reliable decision-making processes. It also facilitates Accountability, a pillar mainly driven by policies over the AI assets that will be transferred to industry.
Therefore, an emerging trend of AI-based Blockchain data processing is paving the way for streamlined integration between these two cutting-edge technologies \cite{salama_ai_2022}.

\subsubsection{Federated Learning}
Building upon the aforementioned insights in previous sections, as depicted in Figure \ref{fig:RQ1_2}, there is a prominent trend where the domains of IoT \& Big Data, and Cloud Computing prominently align with the critical aspects of Privacy and Data Governance requirements. This convergence highlights the substantial emphasis placed on safeguarding privacy and ensuring robust Data Governance, with these aspects constituting a significant proportion of the overall focus with 14.9\% and 10.2\%, respectively. 
A promising approach to enhance privacy in the utilization of AI, particularly in the context of IoT, Cloud Computing, and Edge services, is Federated Learning. 
Federated Learning is an innovative AI technique that allows systems to train models locally, leveraging local updates to collaboratively build a global AI asset without the need to transfer sensitive data externally \cite{lo_towards_2022, yang_federated_2020}. This decentralized approach enables the preservation of data privacy while enabling effective model training and knowledge sharing across distributed systems. 

In Federated Learning, each participating device, such as IoT devices, trains its local model using its own local data, thereby keeping the data within its control and avoiding the need to transmit it externally. Instead, only the locally updated model parameters are communicated to a central server or aggregator, where they are aggregated to construct a global model \cite{lo_towards_2022, yang_federated_2020}. This distributed learning paradigm offers significant advantages in terms of data privacy and security, as it reduces the risk of data exposure and unauthorized access to sensitive information. 

Another option is Split Learning. Split Learning is an emerging distributed ML paradigm that addresses privacy preservation challenge by keeping the data on the client devices \cite{wu2023topology, li2022federated}. In Split Learning, the training process is divided between client devices and a central server. The client devices perform initial computations and transmit the extracted features to the server for further model training. In contrast, Federated Learning involves training models on decentralized devices (such as mobile devices or Edge nodes) without transferring the raw data to a central server. The model is trained locally on each device, and only the model updates are aggregated on the server. Split Learning offers several advantages, including enhanced privacy preservation, reduced communication overhead, and improved scalability. It finds applications in various domains such as healthcare, IoT, and Edge Computing, where data privacy and resource constraints are critical concerns. Privacy-preserving mechanisms, such as differential privacy, can be integrated into Split Learning to provide rigorous privacy guarantees. Secure aggregation protocols are employed to protect the integrity of aggregated model updates. 
In summary, Split Learning offers a mechanism to collectively train models without directly exchanging raw data, preserving the privacy of the contributing parties. The relationship between Split Learning and Federated Learning is also explored to address privacy and security concerns \cite{thapa2022splitfed}.

The integration of Federated Learning and Split Learning with IoT, Cloud, and Edge services offers a compelling solution for privacy-sensitive applications. By keeping data local and performing model updates collaboratively, these techniques enable privacy-preserving AI applications in scenarios where data confidentiality is of utmost importance. This approach facilitates the training of accurate and robust models while ensuring compliance with privacy regulations and addressing concerns related to Data Governance. The implementation of an appropriate distribution policy further enhances the integration of systems and models. For instance, when multiple AI processes are employed on equal operational units under similar operational conditions but in different locations, a suitable distribution policy facilitates the coordination and collaboration of IoT devices. Through this collaborative effort, the sensed information from IoT devices is effectively coordinated and utilized, leading to the development of a shared global model that demonstrates high prediction accuracy. This collaborative learning process can be performed either at the Edge or in the Cloud, depending on the specific requirements and constraints of the application. 

However, it is important to note that collaborative learning techniques also pose several challenges. These include addressing communication and bandwidth limitations, dealing with heterogeneous and unevenly distributed data, ensuring model Fairness and Accountability, and managing the coordination and synchronization of local model updates across devices \cite{lo_towards_2022, yang_federated_2020}. Overcoming these challenges requires careful design and optimization of the learning process, as well as the development of appropriate mechanisms for data aggregation, model evaluation, and privacy preservation. According to Yang et al. \cite{yang_federated_2020}, advanced ML learning techniques in Federated Learning can effectively mitigate learning latency by eliminating the need to offload data to a centralized data center. This approach allows for local model updates on selected devices, thereby minimizing the communication overhead associated with transmitting data to a central server. However, despite the potential benefits, the process of updating the selected devices can pose challenges in terms of communication bottlenecks, particularly in intelligent system applications.

Another challenge is the additional Trustworthy AI risks related to Accountability and Fairness. These risks arise from the involvement of multiple stakeholders and devices, as well as the potential for uneven or heterogeneous distribution of data across these entities. Ensuring Accountability in distributed learning processes requires mechanisms for tracking and verifying the actions and decisions of all participants involved in the collaborative learning process. Similarly, achieving Fairness needs the development of strategies to mitigate biases and ensure equitable representation and treatment of data across different devices and stakeholders. Addressing these Trustworthy AI risks in such environments requires careful consideration and innovative solutions. Future research efforts should focus on developing robust frameworks and methodologies that enhance Accountability and Fairness in Federated Learning while preserving the privacy and security of participant data.

Lo et al. \cite{lo_towards_2022} propose a methodology that combines a Blockchain-based Trustworthy Federated Learning architecture to mitigate the risks associated with these Trustworthy AI considerations. The integration of Blockchain technology provides a transparent and immutable record of participant actions, ensuring Accountability and traceability in the Federated Learning process. This approach, along with other advancements in the field, aims to enhance the Privacy and Data Governance aspects of Federated Learning.

Furthermore, other researchers have also explored the fusion of Blockchain and Federated Learning to promote privacy and Data Governance. Fan et al. \cite{fan_hybrid_2021} propose a hybrid Federated Learning framework that integrates Blockchain technology to ensure secure and privacy-preserving data sharing among participants. Zhang et al. \cite{zhang_blockchain-based_2021} present a Blockchain-based approach for Federated Learning, focusing on data integrity, privacy, and incentive mechanisms to encourage active participation and cooperation among the Federated Learning participants.

These works demonstrate the growing interest in integrating Blockchain and Federated Learning to address Accountability, Fairness, Privacy, and Data Governance concerns in collaborative AI settings. Further research is needed to explore the full potential of these approaches and develop comprehensive frameworks that effectively balance the diverse requirements of Trustworthy AI in Federated Learning scenarios.

In order to establish Accountability in distributed learning, Lo et al. \cite{lo_towards_2022} propose the adoption of a smart contract-based data-model provenance registry. This registry ensures a transparent and immutable record of the data and model updates contributed by each participant in the Federated Learning process. Furthermore, the authors introduce a weighted fair data sampler that promotes Fairness during the training process by addressing the potential bias introduced by varying data distributions among participants. Their approach demonstrates promising results, showing improved generalization and accuracy compared to traditional Federated Learning settings.

\subsubsection{Machine Learning Operations - MLOps}
In addition to the use of smart contracts and fair data sampling, Accountability in Federated Learning can also be enhanced through the adoption of MLOps practices. MLOps encompasses a set of principles and techniques aimed at effectively deploying, managing, and monitoring ML models in production. It involves collaboration between software development (DevOps) and data science teams to ensure reliable and efficient ML model operations.

While the literature on MLOps in the context of Federated Learning is still limited, its potential for enhancing Accountability and Transparency in the involved processes is evident. By defining proper processes and workflows, MLOps can provide an extension to various AI applications, enabling better governance and Accountability throughout the AI life-cycle. Further research and exploration of MLOps are warranted to fully realize its benefits and establish best practices in this evolving field.

Furthermore, Fairness and societal and environmental well-being presents an opportunity for contributions on Trustworthy AI but not on Industry 4.0. In addition to Figure \ref{fig:RQ3_3}, in Figure \ref{fig:RQ1_2} an important gap is shown in topics related to process planning, quality control and predictive maintenance. While in the latter it is complex to find a relation, the other two start presenting some work related to equitable treatment of individuals or groups, ensuring that the benefits and opportunities derived from Industry 4.0 technologies are distributed without bias or discrimination \cite{he2021maintenance}. Fairness plays a pivotal role in addressing societal concerns and promoting inclusivity in the deployment of emerging technologies. By incorporating Fairness principles into the design, development, and implementation of Industry 4.0 systems, equal access can strive for, unbiased decision-making, and the elimination of discriminatory practices. Achieving Fairness in Industry 4.0 requires the identification and mitigation of potential biases in data collection, algorithmic decision-making, and human-computer interactions, that will have an impact on the production systems and quality control processes. It requires sandboxing environments to test the processes in a no-harming manner, followed by robust regulatory frameworks, ethical guidelines, and Accountability mechanisms to ensure that the benefits of industry are shared equitably across diverse populations and that the inclusion of these processes does not impact the ongoing productions. By integrating Fairness as a core principle, industry can contribute to a more just and inclusive society. This can be achieved by exploiting on different phases of development \cite{ijcai2023p735}, and MLOps techniques are a quite useful methodology for understanding the phases where the developers should take care to avoid undesired bias by using sandboxing environments.

\subsubsection{Big Data Stream Processing}
As depicted in Table \ref{table:industry5_facet}, the attributes intrinsic to the domain of Big Data Stream Processing exhibit a direct nexus with the fundamental tenets of Trustworthy AI. 
The integration of Big Data and stream processing with Trustworthy AI has the potential to provide a solid foundation for achieving data privacy and robustness \& safety requirements. The paradigm can effectively address these through the judicious implementation of robust data anonymization, encryption, and access control mechanisms. Privacy-preserving techniques can be implemented during real-time data processing. This integration can help to ensure that personal information remains secure while extracting insights from streaming data, keeping access control mechanisms.

In addition, the Big Data Stream Processing facilitates real-time data pre-processing, cleansing, and filtering processes. 
These processes serve to augment the quality and dependability of the data reservoir exploited by AI systems. It provides expeditious processing, visualization, surveillance, and real-time analysis of streaming data. This covers a pivotal need to endow AI systems with the competence to effectuate prompt and precision-guided decisions and can address human agency and oversight by implementing human-in-the-loop and human-in-control processes, thus enhancing the trustworthiness of AI systems. In addition, with respect to Transparency and Accountability, key requirements to address the black-box nature of AI algorithms and models deployed in industry, Big Data Stream Processing can facilitate the mechanisms to capture and process data in a transparent and auditable manner\cite{sahal2020big}. 

Although Big Data Stream Processing has the potential to contribute significantly to various aspects of Trustworthy AI, the existing literature lacks comprehensive coverage of its impact on ensuring Trustworthy AI. In fact, we could only locate a single manuscript that partially addresses the consideration of this technology's influence on Trustworthy AI, as mentioned in \cite{chahed2023aida}. Despite the potential benefits that the technology has, it focuses on the broader impacts of IoT and Big Data, lacking specific discussions related to Trustworthy AI.


\subsection{Mapping between Industry 4.0 and Industry 5.0}
\begin{figure*}[htbp!]
    \centering
    \includegraphics[width=120mm]{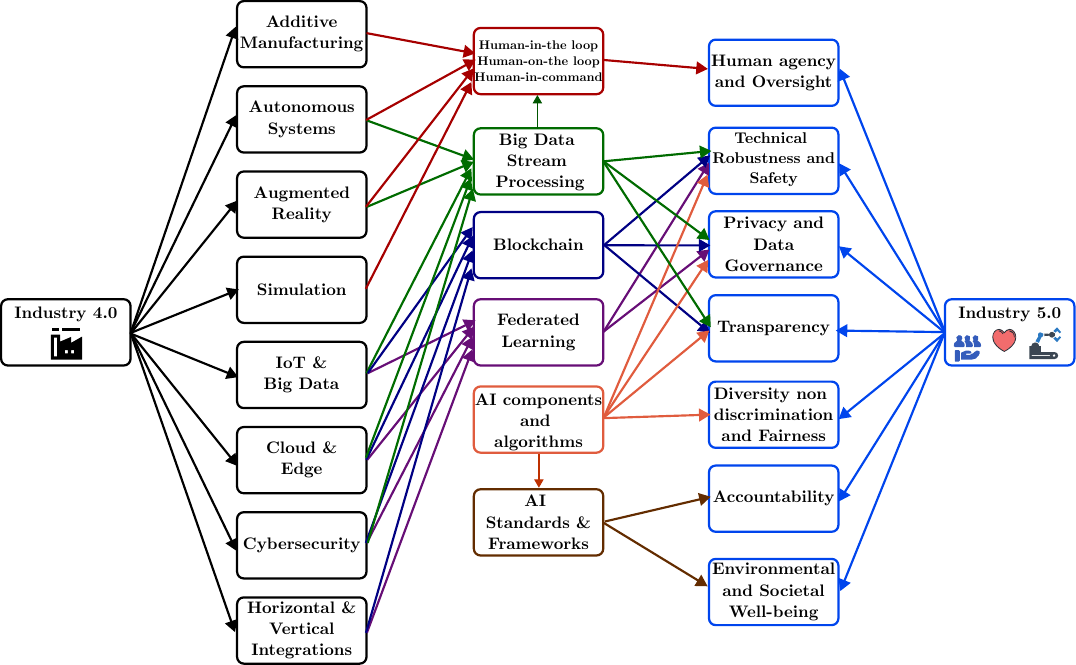}
    \caption{Fostering Industry 5.0 - linkage of technology concepts with new topics and Trustworthy AI requirements}
    \label{fig:Fostering map process}
\end{figure*}

Figure \ref{fig:Fostering map process} illustrates the interconnection between Industry 4.0 and its Technology Enablers with various techniques and concepts that have the potential to contribute to the fulfilment of Trustworthy AI requirements. 

The primary objective of this map is to provide a comprehensive overview rather than offering prescriptive guidance on how to ensure AI trustworthiness. By highlighting the potential connections between Industry 4.0, its Technology Enablers, and various techniques and concepts, this map aims to facilitate a broader understanding of the landscape and stimulate further exploration and research in the pursuit of Trustworthy AI.

On the left side of the diagram, the Industry 4.0 pillars are listed in a column. On the right part, the Trustworthy AI requirements and their linkage to Industry 5.0.

The middle column shows the specific techniques and concepts as Technical components, Algorithms and Standards, and Frameworks, which have a transversal impact on the Technology Enablers of Industry 4.0. These components serve as foundational elements that can be leveraged to address Trustworthy AI challenges.
Blockchain technology, known for its decentralized and immutable nature, enables secure and transparent transactions and data management. By leveraging Blockchain, Industry 4.0 can enhance the traceability and auditability of AI processes, thereby addressing the trustworthiness aspect of AI systems. Similarly, Federated Learning, an AI technique that allows training models locally on individual devices without sharing raw data, preserves data privacy and confidentiality. By adopting Federated Learning, Industry 4.0 systems can achieve higher levels of privacy and data security, mitigating concerns related to data leakage and unauthorized access.

The inclusion of MLOps in the map is limited due to insufficient specification and documentation in the existing literature. However, MLOps encompasses the practices and processes involved in deploying and managing ML models having the potential to be applied to any Industry 4.0 pillar to operationalise these through the use of AI. It enhances Accountability, Transparency, and reproducibility in AI systems. However, further research and development are required to elucidate the specific contributions of MLOps to Trustworthy AI within the context of Industry 4.0.

Big Data Stream Processing has a direct connection with the human-in-the-loop, human-on-the-loop and human-in-command that are enablers for Additive Manufacturing, Autonomous Systems, Simulation and Digital Twins and Augmented Reality. Furthermore, the nature of Big Data Stream Processing also have strong connections with deployment environments, Cloud, Edge and IoT, Big Data, and Cybersecurity. In addition, the enabler for Big Data Stream Processing goes for technical Robustness, Privacy and Data Governance, and Transparency.

The AI components and algorithms, that is core for the drivers, has a strong link with all the components in Industry 4.0. Not depicted on the diagram to avoid an additional crosslink. Furthermore, it provides direct input into AI standards and frameworks needed for Accountability and the legal part of environmental and societal well-being required for Industry 5.0.

\section{Conclusions}
\label{conclusions}
A general perspective of the readiness of Industry 4.0 to Industry 5.0 was analyzed in the present work, examining the advancements, challenges, and emerging trends in the realm of industrial automation and AI. Through the analysis of existing literature, we have gained valuable insights into the key technologies and concepts driving this transition and their implications for the future of industrial systems.

Industry 4.0 has revolutionized traditional manufacturing processes by leveraging technologies such as IoT, Big Data analytics, Cloud Computing, and cyber-physical systems. These advancements have enabled increased automation, connectivity, and data-driven decision-making, leading to enhanced productivity, efficiency, and competitiveness in industrial sectors. However, as we move towards the era of Industry 5.0, new opportunities and challenges arise, necessitating the integration of AI and intelligent systems into the industrial landscape.

The emergence of AI in Industry 5.0 brings about significant transformations in the way industrial processes are managed and optimized. AI technologies such as ML, Deep Learning, and Natural Language Processing empower industrial systems to autonomously learn, adapt, and make intelligent decisions. This paradigm shift towards intelligent and Autonomous Systems has the potential to unlock new levels of efficiency, flexibility, and customization in industrial operations.

This document sheds light on several Technology Enablers that are driving the transition from Industry 4.0 to Industry 5.0. Blockchain, with its decentralized and transparent nature, offers enhanced security, data integrity, and privacy for industrial systems. Federated Learning, as an AI technique, enables collaborative and privacy-preserving model training across distributed IoT devices, ensuring data privacy and security 
Big Data processing is a key enabler for Transparency, Robustness, Privacy and Data Governance by enabling the fast process of data to be comprehensively visualised by humans. These technologies, along with others like MLOps, are poised to play a pivotal role in fostering Trustworthy AI and addressing the challenges associated with Accountability, Fairness, and Transparency in industrial systems.

Furthermore, it was revealed the importance of considering the ethical, legal, and societal implications of the transition to Industry 5.0. As AI becomes increasingly embedded in industrial processes, it is crucial to ensure responsible and ethical deployment, taking into account issues such as bias, Fairness, and human-machine collaboration. Additionally, the regulatory frameworks and standards governing AI in industrial settings need to evolve to address the unique challenges and opportunities presented by Industry 5.0.

In conclusion, the transition from Industry 4.0 to Industry 5.0 represents a significant evolution in industrial automation, driven by the convergence of AI, IoT, and advanced analytics. The study shows that the most developed Trustworthy AI requirements are directly linked to technical components. At the same time, the lowest is not directly dependent on the industrial sector since they require a clear definition of protocols, regulations, and implementation approaches (standards) that facilitate the transition towards Industry 5.0. Finally, attention was driven towards technological approaches such as Blockchain, Federated Learning, MLOps, and ontological approaches to foster different Trustworthy AI requirements. This survey has provided an overview of the current state of the field, highlighting the key technologies, challenges, and potential solutions in the journey towards Industry 5.0. As we navigate this transformative phase, it is crucial for researchers, industry practitioners, and policymakers to collaborate and address the technical, ethical, and societal considerations to ensure the successful and responsible integration of AI in the industrial landscape. By embracing these advancements and leveraging the potential of Industry 5.0, we can unlock unprecedented levels of productivity, sustainability, and innovation in the future of manufacturing and beyond.

\section{Acknowledgements}
This paper has been partially supported by the European Commission by funding the ASSISTANT project (no. 101000165), AI4Europe project (no 101070000) and the Science Foundation Ireland under Grant No. 12/RC/2289-P2 for funding the Insight Centre of Data Analytics, co-funded under the European Regional Development Fund. We would like to personally thank Mr. Shaun Gavigan for the proofreading of this paper.

\bibliographystyle{ACM-Reference-Format}
\bibliography{ai-ethics-man}

\end{document}